\documentclass[lettersize, journal]{IEEEtran}
\usepackage{amsmath,amsfonts}
\pdfoutput=1

\usepackage{algorithmic}
\usepackage{algorithm}
\usepackage{array}
 \usepackage{color}
\usepackage[caption=false,font=normalsize,labelfont=sf,textfont=sf]{subfig}
\usepackage{textcomp}
\usepackage{stfloats}

\usepackage{verbatim}
\usepackage{graphicx}
\usepackage{cite}
\usepackage{soul}
\usepackage{url}

\def\cG{\mathcal{G}}

\def\cV{\mathcal{V}}
\def\cW{\mathcal{W}}

\def\bX{\mathbf{X}}
\def\bx{\mathbf{x}}
\def\bu{\mathbf{u}}

\begin{document}

\title{Analysis of the Spatio-temporal Dynamics of COVID-19 in Massachusetts via Spectral Graph Wavelet Theory}
\author{Ru~Geng,
        Yixian~Gao,
        Hongkun~Zhang,
        and~Jian~Zu       
\thanks{Manuscript received...

YX Gao is partially supported by NSFC grants (11871140, 12071065)
and National Key R\&D Program of China (2020YFA0714102). HK Zhang is partially supported by Simons Foundation Collaboration Grants for Mathematicians (706383). J Zu is partially supported by NSFC grants (11971096, 11971095).(Corresponding author: Jian Zu .)}
\IEEEcompsocitemizethanks{\IEEEcompsocthanksitem Ru~Geng, Yixian~Gao and Jian~Zu are with Center for Mathematics and Interdisciplinary Sciences, and School of Mathematics and Statistics, Northeast Normal University, Changchun, 130024, P.R. China (e-mail: zuj100@nenu.edu.cn).

Hongkun~Zhang is with Department of Mathematics and Statistics, University of Massachusetts, Amherst MA 01003, US (e-mail: hongkun@math.umass.edu).\protect\\}
}

\markboth{Journal of \LaTeX\ Class Files,~Vol.~**, No.~*, ****~****}%
{Shell \MakeLowercase{\textit{et al.}}: A Sample Article Using IEEEtran.cls for IEEE Journals}


\maketitle

\begin{abstract} 
The rapid spread of COVID-19 disease has had a significant impact on the world.  In this paper, we study COVID-19 data interpretation and visualization using open-data sources for 351 cities and towns in Massachusetts from December 6, 2020 to September 25, 2021. Because cities are embedded in rather complex transportation networks, we construct the spatio-temporal dynamic graph model, in which the graph attention neural network is utilized as a deep learning method to learn the pandemic transition probability among major cities in Massachusetts. Using the spectral graph wavelet transform (SGWT), we process the COVID-19 data on the dynamic graph, which enables us to design effective tools to analyze and detect spatio-temporal patterns in the pandemic spreading. 
We design a new node classification method, which effectively identifies the anomaly cities based on spectral graph wavelet coefficients.
It can assist administrations or public health organizations in monitoring the spread of the pandemic and developing preventive measures. 
Unlike most work focusing on the evolution of confirmed cases over time, we focus on the spatio-temporal patterns of pandemic evolution among cities.
Through the data analysis and visualization, a better understanding of the epidemiological development at the city level is obtained and can be helpful with city-specific surveillance.
\end{abstract}


\section{Introduction}
\label{sec:introduction}
\IEEEPARstart{W}{hen} writing this paper, the COVID-19 pandemic is still ongoing and has resulted in more than 260 million people diagnosed and more than 5 million deaths worldwide. The COVID-19 spreading was recognized by the World Health Organization (WHO) as a pandemic on March 11, 2020 \cite{WHO2020}. As of December 1, 2021, confirmed cases in the United States surpassed 4,810,000 and Massachusetts surpassed 919,000 \cite{JHU2021}.  
With little information on similar past pandemics, collecting mobility, safety, and behavior data related to COVID-19 and learning from the collected data become the key for decision-makers. The final scale of the disaster had not yet been determined. To make matters worse, the mutation of the virus has led to waves of pandemics, and humankind still faces significant challenges. 

Traditional mathematical models use compartmental models to study the transmission dynamics of COVID-19, such as SIR and SEIR models and their variants, see Gao et al. \cite{gao2021transmission}, Church \cite{church2021}, Neves and Guerrero \cite{neves2020}, Ng and Gui \cite{ng2020}. Miranda et al. \cite{miranda2021} construct a hybrid ODE-network model for the COVID-19 pandemic accounting for certain spatial aspects. Some state-of-the-art technologies, such as machine learning, and deep learning, have been introduced in the research of COVID-19. For example, Tang et al. \cite{tang2021} study the interplay of demographic variables and social distancing scores in the deep prediction of COVID-19 cases in the United States. Melin et al. \cite{Melin2020} use a self-organizing mapping neural network to cluster countries with similar pandemics so that similar strategies can be used to deal with the spread of the virus.  Tat Dat et al. \cite{TatDat2020} apply wavelet theory and machine learning method to study the evolution of the pandemic in France, Germany, Italy, the Czech Republic, and the US federal states of New York and Florida. Graphs and networks are used to model many real-world problems due to their flexible structure. Li and Mateos \cite{LiMateos2021} conduct a graph Fourier frequency analysis to investigate the county-level contagion patterns of COVID-19. Gao et al. \cite{gao2021} utilize a new graph neural network -- STAN to predict both state-level and county-level future number of infected cases.  Kapoor et al. \cite{kapoor2020} examine COVID-19 forecasting using spatio-temporal graph neural networks. However, there is not much work analyzing the spatio-temporal pandemic spread patterns at the city levels. 

In this paper, we focus on analyzing the local pandemic spread patterns in the State of Massachusetts (MA), USA, which covers a relatively small area of 21,000 square kilometers. Cars are the primary transportation tools among cities and towns in the state of MA. Therefore, the local spread of the pandemic is mainly affected by the transportation graph structure of the state.
Unlike the state-level or county-level graphs mentioned above, our city-level graph is more suitable for the local epidemiological analysis in MA.  In this Massachusetts Route Graph $G_{MR}=(V, E)$, hand-crafted from US Routes, we map cities to vertices on the graph, such that each vertex $\tau_i\in V$ is associated with a set of time-dependent features extracted from biweekly COVID-19 data. 
The edge between two vertices represents that there is a US Route connecting the corresponding cities.
The pandemic spreading between two cities may be affected not only by population movements such as transportation, commuting to work, and tourism; but also by hidden complex factors such as the circulation of COVID-19 virus-contaminated goods on the packaging, etc. Therefore, the traditional method of calculating the edge weight using distance and populations cannot accurately represent the pandemic spread among cities.  
 
\begin{figure*}[htbp]
\centering
\includegraphics[width=17cm]{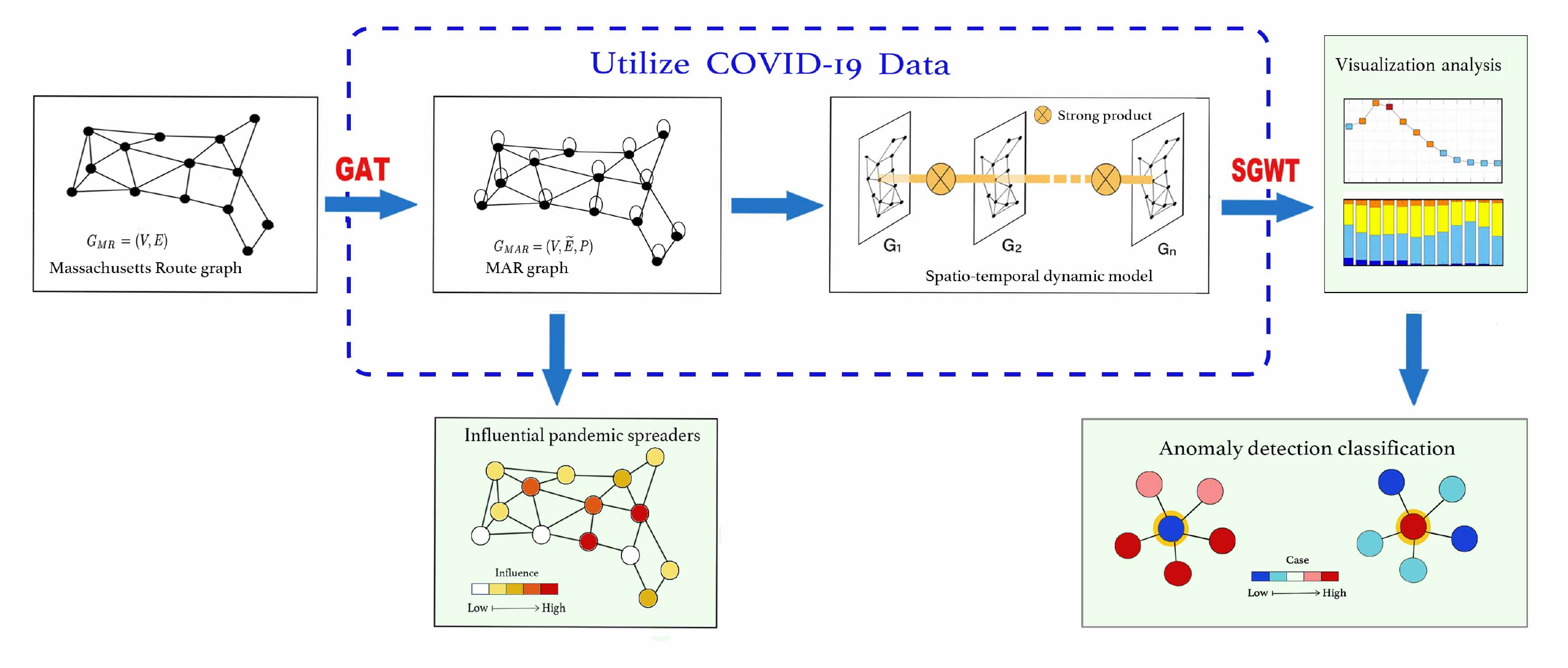}
\caption{The overall architecture of this paper.}
\label{schematic_diagram}
\end{figure*} 
 
This paper further utilizes a deep learning method via graph neural networks -- Graph Attention Network (GAT) -- to learn the Markov transition matrix $P$, in order to capture geographical proximity and data features similarity in the transition probabilities. The resulting graph $G_{MAR}=(V,\widetilde{E},P)$ is called the Massachusetts Attention Route (MAR) graph. GAT has been proved to be efficient for learning the edge weights by adaptive assigning different importance to different nodes through the learning of downstream tasks and data\cite{Velikovi2018, GuGao2019, tang2020, zhou2021, zhang2020, vaswani2017attention}. We also ranked the top influential cities for pandemic spread based on the attention coefficients learned from the GAT. Moreover, we construct the spatio-temporal dynamic graph model connected by the strong product\cite{2011Handbook}.

To analyze the pandemic signals on the spatio-temporal dynamic graph model, we use a powerful tool -- the spectral graph wavelet transform (SGWT). Indeed classical wavelet transformation has played an important role in multiresolution analysis, especially for studying signals containing discontinuities and sharp spikes. However, the construction of wavelets on graphs is rather difficult. Hammond et al. \cite{Hammond2011} and Shuman et al. \cite{Shuman2015} construct SGWT, and successfully apply it in studying graph signals and identifying anomalies. The SGWT coefficients contain rich information about the graph signal; however, it remains a big challenge to interpret them properly for non-experts. In this paper, we use a visualization methodology that relies on SGWT and the structure of our graph to enable the visual multiresolution analysis of time-varying COVID-19 data in Massachusetts. We analyze the dynamic patterns of pandemic signals for cities and towns, and provide visualization diagrams based on our spectral graph wavelet analysis results. Moreover, we define an average metric function based on wavelet coefficients to give cities a spatio-temporal ranking synthetically. The top five ranked cities include Springfield, Amherst, Great Barrington, Holyoke, and Sandisfield, all in line with evidence of repeated poor responses to the pandemic, see \cite{Springfield1, umass3, holyoke}. For example, Amherst has reported pandemic peaks during the starting period of semesters of University of Massachusetts (UMass) Amherst, while Holyoke has reported pandemic breakout in its nursing homes. These all make the pandemic situation more severe than those in the surrounding areas. On the other hand, according to our ranking, we identify the five best-performance cities -- New Marlborough, Harvard, West Tisbury, Tolland, and Leverett. These are in line with evidence showing their excellent pandemic prevention and control work, \cite{marlborough2,marlborough1, Harvard, Harvard3}, which shows that the search for ambassadors at New Marlborough and the offer of free safety training courses in Harvard, make their pandemic situations much better than that of surrounding cities.

Experts often say that determining which cities and states have had the best response to COVID-19 thus far is tough and unfair \cite{news2020}. In this paper we hope our study provides a useful evaluation method. The basic framework and main results of this paper are shown in Fig. \ref{schematic_diagram}.

In summary, the main contributions of this paper include:

 \begin{itemize}
\item Massachusetts Route graph is constructed by connecting cities and towns based on US Routes or interstate highways.
\item GAT learns the weights of the MAR graph, which helps calculate each node's vulnerability affected by the COVID-19 infection rate of surrounding nodes.
\item SGWT is used to analyze the spatio-temporal dynamic model constructed by the strong product.
\item We design a new node classification method, which effectively identifies the anomaly cities based on spectral graph wavelet coefficients.
\end{itemize}

The outline of this paper is as follows: In section \ref{section_2}, we describe the COVID-19 data of cities and towns in Massachusetts. In section \ref{section_3}, we construct the Massachusetts Route graph and learn the pandemic transition probability Matrix by graph attention neural network, then build the spatio-temporal dynamic model of confirmed cases in Massachusetts. Section \ref{section_4} introduces the SGWT and develops the node classification method based on SGWT. In section \ref{section_5}, we visually analyze the overall trend of the pandemic by SGWT, make a refined node classification in anomaly cities, and evaluate the city's pandemic prevention work. In addition, we identify the top influential pandemic spreader in the MAR graph. In section \ref{section_6}, we close with a conclusion section.

\section{COVID-19 Data description}
\label{section_2}
\begin{figure}[htbp]
\centering
\includegraphics[width=8cm]{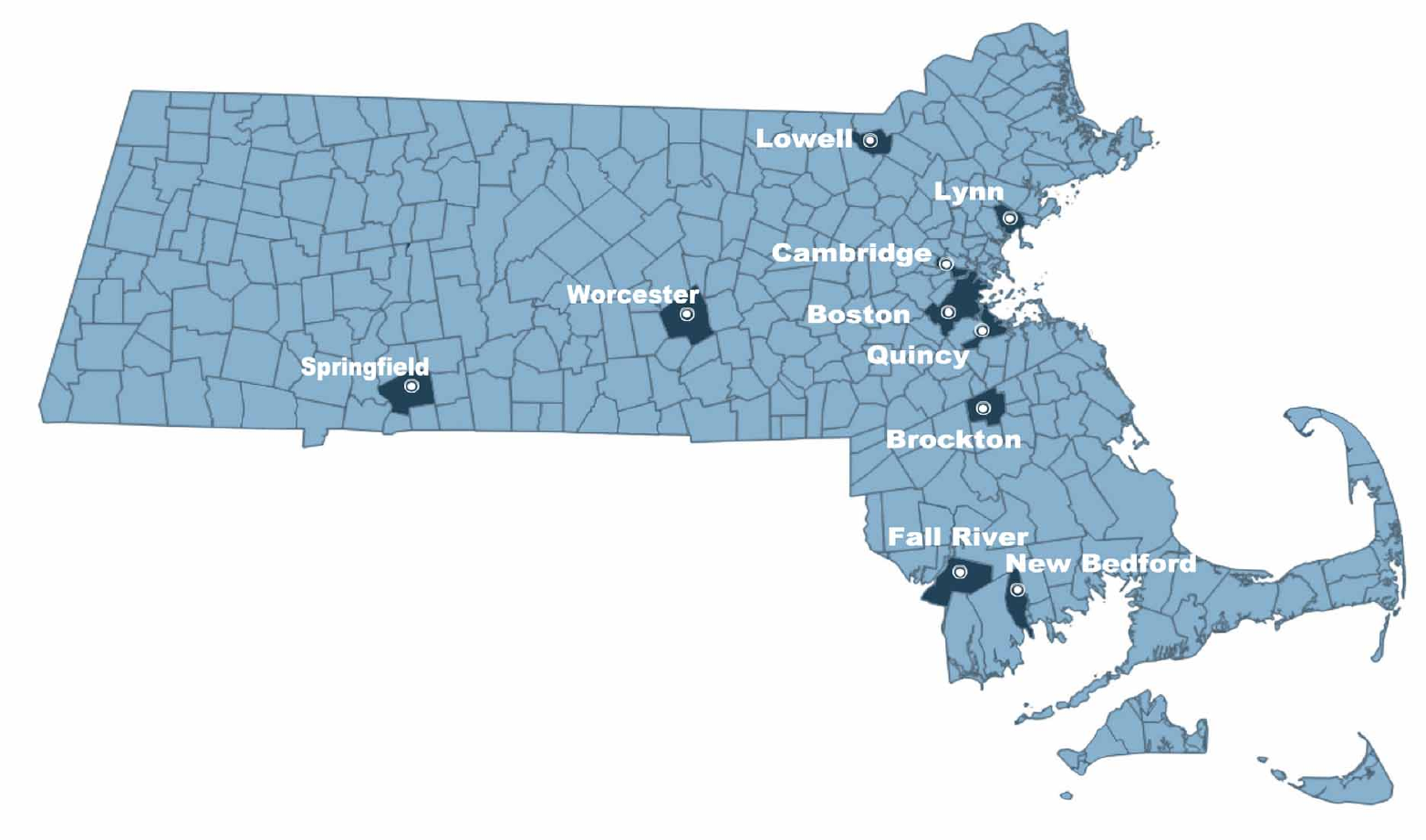}
\caption{Map of Massachusetts.}
\label{ma_map}
\end{figure}

This paper studies the COVID-19 pandemic spread in the state of Massachusetts, USA, which consists of 351 cities and towns. In the MA, the list of 10 largest cities (in population) is Boston (692,958), Worcester (191,575), Springfield (156,245), Lowell (116,143), Cambridge (111,989), Quincy (101,531), Lynn (100,824), New Bedford (99,980), Brockton (99,226) and Fall River (89,317). See Fig. \ref{ma_map} for the locations of these cities.

The dataset available in this paper is composed of population data $\{N_i, i=1,\cdots, 351\}$ and the "two-week-case" time series data $\{\gamma_i(t), t=1,\cdots, 41\}$ at the city/town level in Massachusetts, from December 6, 2020 to September 25, 2021, for 41 weeks total, which is collected from the official website \cite{massdata}.
We define the biweekly confirmed COVID-19 cases per thousand population (abbreviated as confirmed cases), denoted by the multivariate time series 
\begin{equation}\label{graphsig}
\mathbf{x}_i (t) = 1000 * \frac{\gamma_i(t)}{N_i},\,\,\,\,\, i=1,\cdots, N,\ t=1,\cdots, T
\end{equation}
with $N=351$ and $T=41$. 

The multivariate time series data $\bx_i(t)$ will first be used to learn the pandemic transition probability $p_{i,j}$ from city $i$ to city $j$, in order to better capture the spatio-temporal evolution of the local pandemic spread dynamics. After that, their evolution patterns will be analyzed using the spatio-temporal dynamic graph model, and we are able to identify the most influential spreaders (cities), and detect anomaly cities with unusual pandemic spread patterns. 

\section{Massachusetts Attention Route Graph and spatio-temporal dynamic model}
\label{section_3}
In this section, we first construct the Massachusetts Route graph $G_{MR}=(V,E)$ based on US Routes, then learn the attention graph weight $P$ from the COVID-19 data using GAT to get the Massachusetts Attention Route (MAR) Graph $G_{MAR}=(V,\widetilde{E},P)$. The strong product is utilized to construct a spatio-temporal dynamic graph model. 
\subsection{Construction of the Massachusetts Route graph}
We take the longitude and latitude coordinates of the cities and towns as the vertices $\tau_i\in V$ from the website \cite{toolforge}. 
The detailed list of cities and their corresponding vertex ID is included in the Appendix.
An edge $e_{i,j}\in E$ represents the existing US Routes or interstate highways go through the administrative regions of the cities $\tau_i, \tau_j\in V$ successively.
We also connect the cities of Barnstable (No.21), Nantucket (No.197), and Oak Bluffs (No.221) because of the heavy sea transportation among them.
The number of cities and towns in our study is $|V|=N=338$\footnote{There are 13 small towns located in rugged areas without any highway or heavy sea transportation. We do not consider them in our following study. }. 
Let $A\in\mathbb{R}^{N\times N}$ be the adjacency matrix of the graph $G_{MR}$, such that $a_{i,j}=1$ if $e_{i,j}\in E$, and $a_{i,j}=0$ otherwise. 
Each vertex has signals $\mathbf{x}_i \in \mathbb{R} ^{T}$, $\mathbf{\gamma}_i \in \mathbb{R} ^{T}$ and $N_i \in \mathbb{R} $ defined as in formula (\ref{graphsig}).

\begin{figure}[htbp]
\centering
\includegraphics[width=9cm,height=5cm]{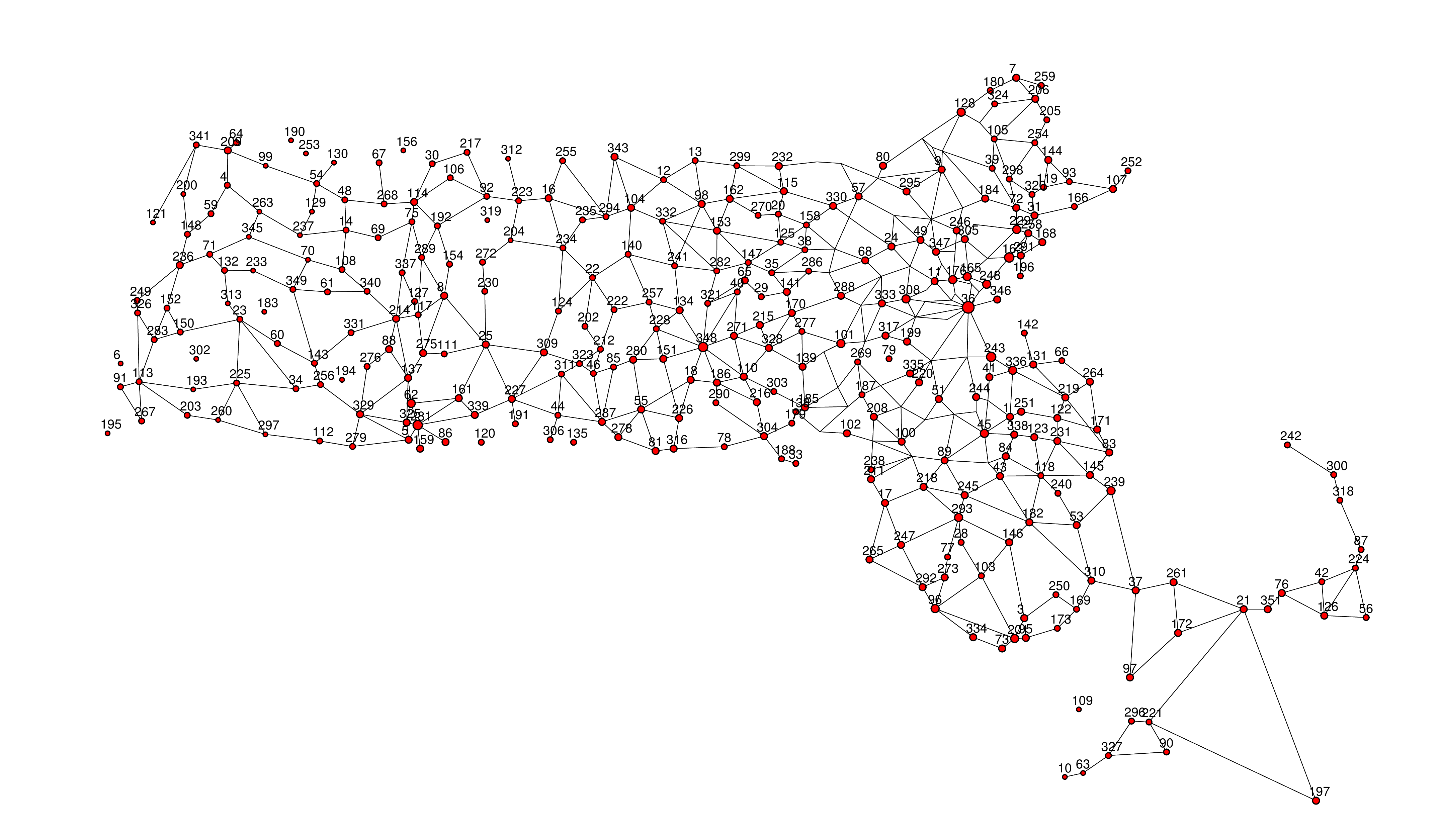}
\caption{Massachusetts Route graph $G_{MR}$.}
\label{map}
\end{figure}

The size of vertices in Fig. \ref{map} is proportional to the city population. For example, Vertex No. 36 represents metropolitan Boston, with the largest population in Massachusetts.

Since the cities within the range between 71$^{\circ}$E to 71.5$^{\circ}$E and 42$^{\circ}$N to 42.8$^{\circ}$N are dense, to improve the visualization in Fig. \ref{map}, we use the downsampling method \cite{vaishnav2016graph} by hiding some nodes in the Massachusetts Route graph. Specifically, we hide the nodes with negative signs in the nodal domain of the largest eigenvector of the Laplacian of the Massachusetts Route graph $G_{MR}$. 


\subsection{Massachusetts Attention Route graph learned by graph neural network}

The pandemic spreading between two cities may be affected by many complex factors, including transportation, tourism, commuting, and other hidden factors such as the circulation of COVID-19 virus-contaminated goods.
Due to the complexity of the pandemic, the traditional weighting method by distance is not a good choice in modeling the pandemic spread between cities (towns). In order to accurately capture the pandemic transition probabilities between cities, we utilize a deep learning method via GAT, to construct the transition probability matrix $P=\{p_{ij}\}$, based on the importance of geographical proximity on the Massachusetts Route graph $G_{MR}=(V,E)$ and data feature similarity. 
The resulting new directed graph with self-loops is called the Massachusetts Attention Route (MAR) graph, denoted as $G_{MAR}=(V,\widetilde{E},P)$. The edge set $\widetilde{E}$ includes $E$, as well as self-loops.

GAT is a powerful deep learning method proposed in \cite{Velikovi2018}, which has state-of-the-art performance on node classification and link predictions. Instead of a simple mean aggregation, the GAT learns to use weighted summation to aggregate the features of neighboring nodes to update the features of the nodes, and automatically adjust the weight through downstream tasks \cite{Velikovi2018,GuGao2019,tang2020,zhou2021,zhang2020}.
Given $N$ multivariate time series $\textbf{x}_i(t)$, for $i=1,\cdots, N$, $t=1,\cdots, T$. Then $\textbf{x}_i\in \mathbb{R}^{T}$ can be viewed as the $T$-dimensional input features of the node $\tau_i$.
We use edge classification as the downstream task, for our GAT neural network, and use the normalized attention matrix in the last layer of the neural network as our Markov transition matrix $P= \{p_{ij}\}$.

\begin{figure*}[h]
\centering
\includegraphics[width = 17cm]{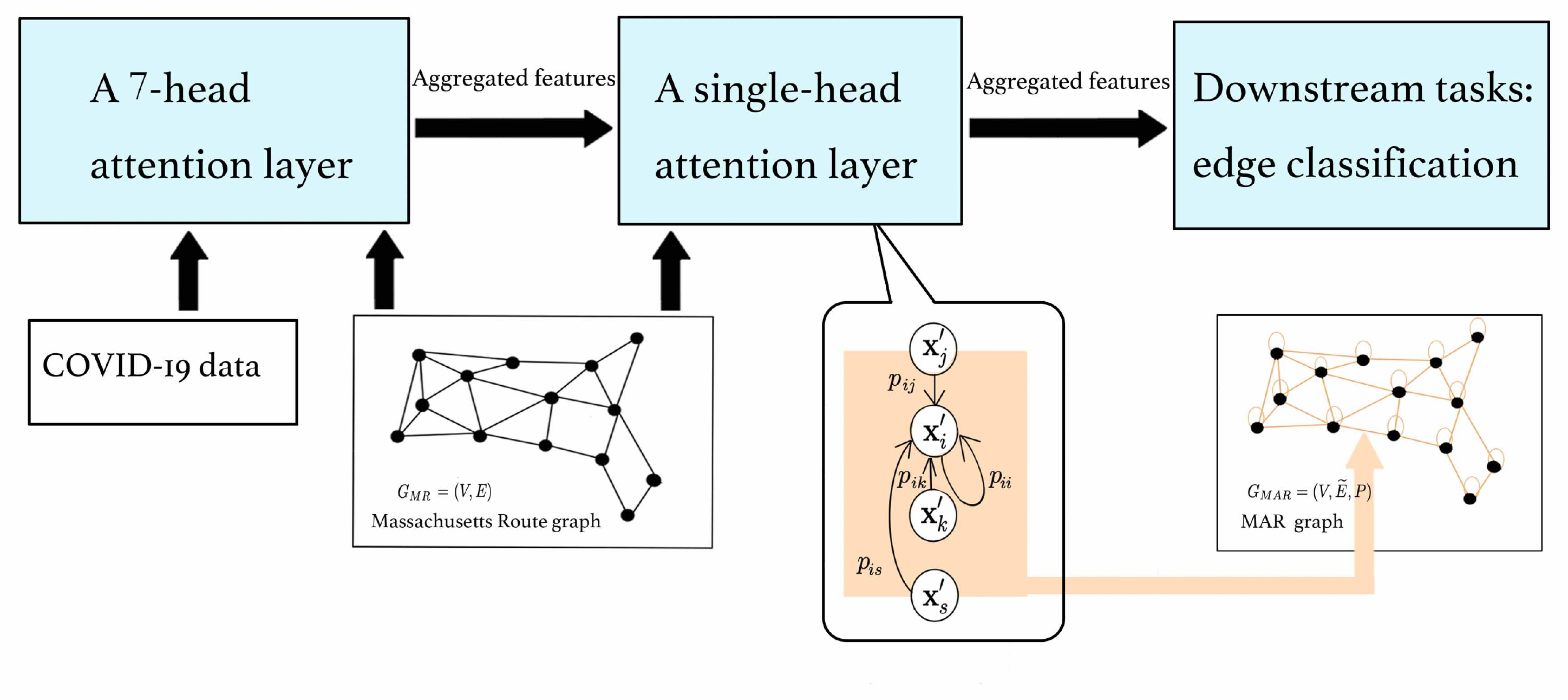}
\caption{GAT framework with edge classification as a downstream task. There are two GAT layers. The first layer is a $7$-head attention layer, and the second layer is a single-head attention layer. The attention coefficient learned by the GAT neural network is applied to define the MAR graph.}
\label{graph_attention}
\end{figure*} 

To improve the model capacity and stabilize the learning process, GAT uses multi-head attention.
Consider a multi-head GAT layer with $K$ heads.
$\textbf{W}^{k} \in \mathbb{R}^{O_k \times T}$, $k = \{1,\cdots, K\}$, is a learnable linear transformation, with $O_k\geq 1$. Let $\hat{\textbf{a}}^{k}\in \mathbb{R}^{2O_k}$ be the $k$-th attention weight vector. The normalized attention coefficients $\alpha_{i,j}^{k}$ capture the importance of $\tau_j$’s features to $\tau_i$'s, expressed as
\begin{equation}\label{atten_coe}
{\alpha}_{ij}^{k}=
\frac{\exp\left(\mathrm{LeakyReLU}(\hat{\textbf{a}}^{k}\cdot[\textbf{W}^{k}\mathbf{x}_i\parallel \textbf{W}^{k}\textbf{x}_j])\right)}
{\sum_{l\in \mathcal{N}_i} \exp\left(\mathrm{LeakyReLU}(\hat{\textbf{a}}^{k}\cdot[\textbf{W}^{k}\mathbf{x}_i\parallel \textbf{W}^{k}\textbf{x}_l])\right) },
\end{equation}
where $\mathcal{N}_i$ denotes the set of first-order neighbours of node $\tau_i$ (including $\tau_i$) and $\parallel$ denotes the concatenation operation, and the activation function is LeakyReLU
\footnote{$
\mathrm{LeakyReLU}(x)=0.35x(1-\mathrm{sgn}(x))/2+x(1+\mathrm{sgn}(x))/2.
$}. 

Let $\textbf{x}_i^{\prime}\in \mathbb{R}^{O}$ be $O$-dimensional output features satisfying $O = \sum_{k=1}^K O_k$.
A $K$-head GAT layer aggregates nodes features across neighborhoods by
\begin{equation*}
\textbf{x}_i^{\prime}=||_{k\in\{1,\cdots,K\}}\mathrm{ELU}\left(\sum_{j\in \mathcal{N}_i} {\alpha}_{ij}^{k} \textbf{W}^{k} \textbf{x}_j\right).
\end{equation*}
where ELU is an exponential linear unit
\footnote{$
\mathrm{ELU}(x)=(\exp(x)-1)(1-\mathrm{sgn}(x))/2+x(1+\mathrm{sgn}(x))/2.
$}.

To learn the attention matrix, we use edge classification \cite{GuGao2019} as a downstream task. 
Let $$M=\{(\tau_i,\tau_j)\in V\times V\,|\, a_{i,j}=1\}$$ be the positive sample set, 
where $a_{i,j}$ is the element of the adjacency matrix $A$ of the graph $G_{MR}$. For any integer $l\geq 1$, the power of the adjacency matrix $A^{l}=(a_{i,j}^{(l)})_{N\times N}$ can be used to define the $l$-th order neighborhood of a vertex.
Let
$$S^{-}=\{(\tau_i,\tau_j)\in V\times V\,|\, a_{i,j}=0, a_{i,j}^{(l)}>0, \exists\,l\in \{2,3\}\}$$ be the negative candidates set. 
The negative sample set $M^{-}$ is randomly generated from the negative instances in $S^{-}$ with a similar size to $M$. 
The GAT neural network is trained by minimizing the following cost function:
\begin{equation*}
\mathcal{L}=-\frac{\sum_{(i,j)\in M \cup M^{-}}
(a_{i,j}\log q_{ij} + (1-a_{i,j})\log(1 - q_{ij}))}{| M\cup M^{-}|}
\end{equation*}
Here $q_{ij}$ is the probability of whether there is an edge between $\tau_i$ and $\tau_j$,
\begin{equation*}
 q_{ij}=\frac{1}{1+\exp\left(-(\textbf{x}_i^{\prime\prime} \odot \textbf{x}_j^{\prime\prime})\cdot\theta\right)},
\end{equation*}
where $\odot$ denotes Hadamard product, $\textbf{x}_i^{\prime\prime}$ is the output feature of the last layer of GAT. $\theta$ is a learned parameter vector and the dimension is the same as $\textbf{x}_i^{\prime\prime}$.

In our setup, there are two GAT layers. Each layer takes an appropriate output dimensionality to improve the performance of the model\cite{nguyen2020wide}. The first layer is a $K$-head attention layer with $K=7$, and the input feature is the graph signal $\mathbf{x}_i\in \mathbb{R}^T$ with $T=41$. The output feature $\textbf{x}_i'$ is cascaded as $O=854$.
i.e. we set $O_k=122$ features each (for a total of 854 features).
The second layer is a single-head attention layer with the $88$-dimensional output features $\textbf{x}_i^{\prime\prime}$.
During the training phase, we divide the sample set $M \cup M^{-}$ into training set, validation set and test set according to the ratio of $6:2:2$.
We also make sure to keep the same number of positive and negative samples in the test set.
The learning rate is set as $lr=0.005$. The early stopping strategy is applied to the validation set to avoid overfitting, with the patience set to 100 epochs.
{\bf The accuracy of edge classification of $G_{MR}$ without self-loops in our result is 0.9499}.

The transition probability $p_{ij}$ is defined as the GAT coefficients of the last GAT layer, i.e.
\begin{equation*}
p_{ij}=
\frac{\exp\left(\mathrm{LeakyReLU}(\hat{\textbf{a}}\cdot[\textbf{W}\mathbf{x}_i^{\prime}\parallel \textbf{W}\textbf{x}_j^{\prime}])\right)}
{\sum_{l\in \mathcal{N}_i} \exp\left(\mathrm{LeakyReLU}(\hat{\textbf{a}}\cdot[\textbf{W}\mathbf{x}_i^{\prime}\parallel \textbf{W}\textbf{x}_l^{\prime}])\right) },
\end{equation*}
where $\hat{\textbf{a}}$ and $\textbf{W}$ defined above are the learned attention weight vector and the learned parameter matrix of this single-head attention layer. One advantage is that the self-attention coefficient $p_{i,i}>0$, which also explains the self-influence of the pandemic in a city. Consequently, the obtained MAR graph is a graph with self-loops.
Fig. \ref{graph_attention} shows the basic framework of this method.


\subsection{Spatio-temporal dynamic graph model}

\begin{figure}[htbp]
\centering
\includegraphics[width=8.5cm]{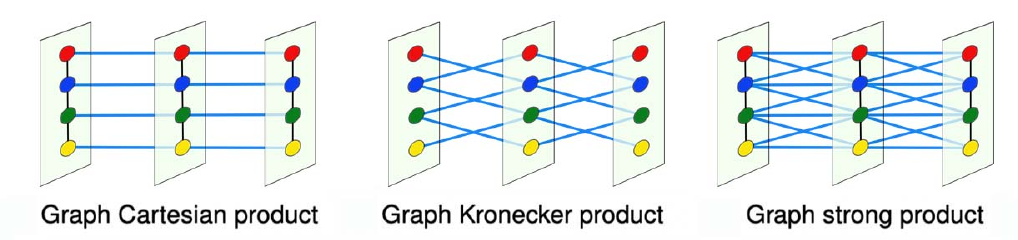}
\caption{Graph products of a spatial path graph (of 4 nodes) and a temporal path graph (of length 3).}
\label{group_product}
\end{figure}

To perform the spatio-temporal node classification using the time-dependent data $\{\textbf{x}_i(t), i=1,\cdots, N,\, t=1,\cdots, T\}$ defined in (\ref{graphsig}), we introduce the concept of graph product. The three well-known graph products are the graph Cartesian product, graph Kronecker product, and graph strong product, see Fig. \ref{group_product}. These have been powerful in studying time-vary data on graphs, see \cite{Sandryhaila2014,Valdivia2015,DalCol2017,DalCol2018}.

The graph Cartesian product only adds extra edges between nodes on the adjacent graph time slices in series, which often weakens the influence of spatial neighbors in the study of SGWT for graph signals. Here we adapt the graph strong product, which also adds temporal edges between pairs of one-hop neighbors in its adjacent graph time slice.

\begin{figure}[h]
\centering
\includegraphics[width=8.5cm,height=5cm]{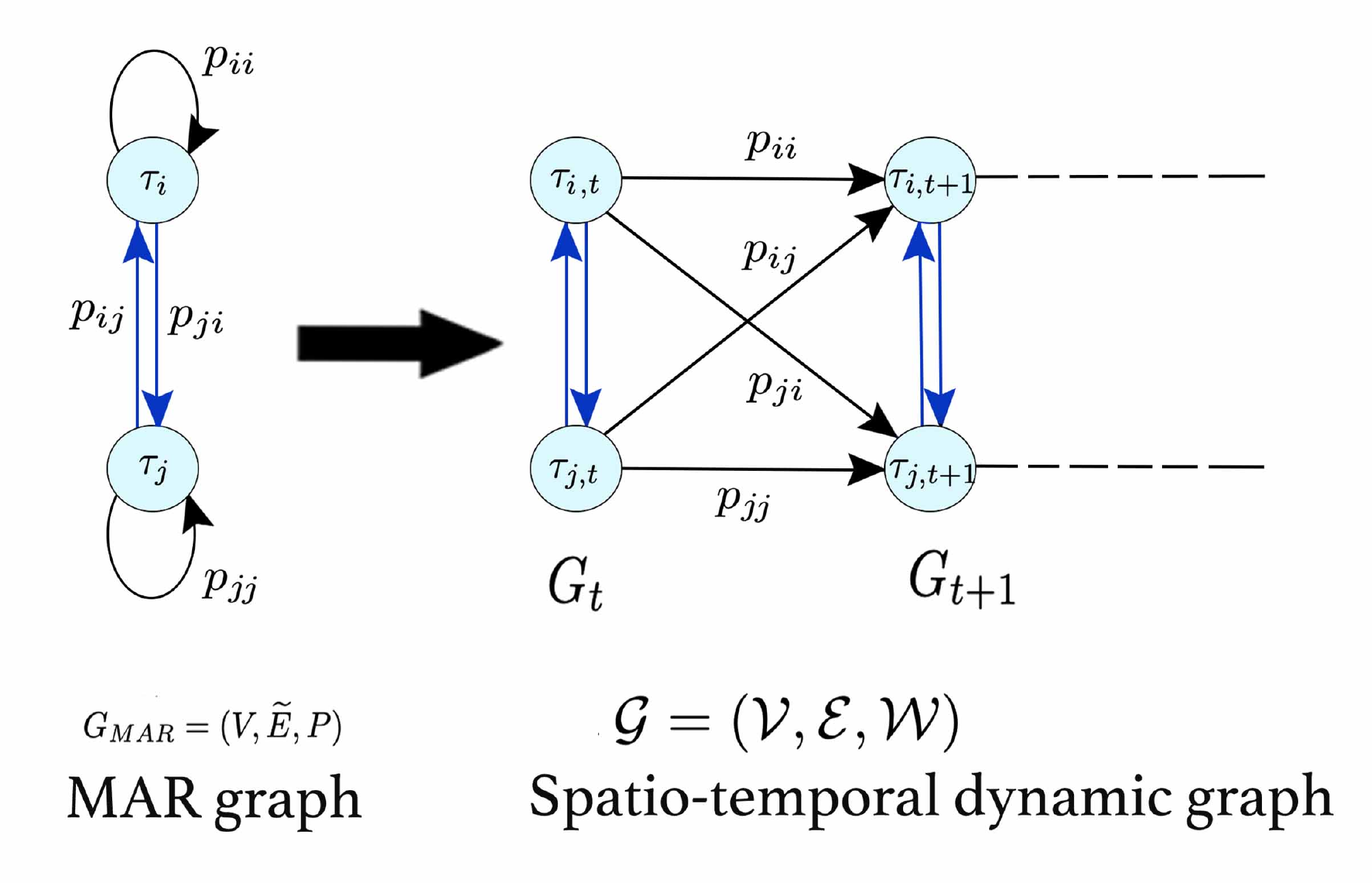}
\caption{Construction of spatio-temporal dynamic graph based on MAR graph. The black edge in the spatio-temporal dynamic graph represents the temporal edge, and the blue edge is the spatial edge inherited from the MAR graph without the self-loop. }
\label{dynamic_product}
\end{figure}

Let $H$ be a path graph consisting of $T$ nodes. The strong product graph can be defined by the product of the Massachusetts Route graph $G_{MR}$ and the path graph $H$. We assign weights to the strong product graph based on the transition probability matrix $P$ of $G_{MAR}$. The resulting directed graph $\mathcal{G}$ without self-loops is the spatio-temporal dynamic model. More precisely, the graph time slices $\{G_t=(V_t,E_t, P_t), t=1,\cdots, T\}$ are copies of graph $G_{MR}$ by assigning weights $P_t=P-\mathrm{diag}(p_{11},p_{22},\cdots,p_{NN})$.
The direction of the temporal edge (black edge in Fig. \ref{dynamic_product}) is from $G_{t}$ to $G_{t+1}$, which represents the time-evolution of the pandemic among these cities.
The weights of the new temporal edges are carried from $P$. Now, the strong product graph can be represented as
$\mathcal{G}=(\mathcal{V}, \mathcal{E}, \mathcal{W})$ with 
$\mathcal{V}=\{\tau_{i,t}, i=1,\cdots, N;\, t=1,\cdots, T\}$, and $\mathcal{E}=(E_1\cup \cdots\cup E_T)\cup (E_{1,2} \cup \cdots \cup E_{T-1,T})$,
where $E_{t,t+1}$ is the collection of temporal edges connecting $G_t$ and $G_{t+1}$.
The new weight matrix $\cW$ is not normalized anymore, as we have added extra temporal edges, which carry the same weight as the spatial edges. More precisely, the edge $(\tau_{i,t},\ \tau_{j,t+1})$ has the same weight $p_{ji}$ as the spatial edge $(\tau_{i,t},\ \tau_{j,t})$, see Fig. \ref{dynamic_product}. Note that $\mathcal W$ is rather sparse, as we only added temporal edges between two adjacent graph time slices.

\section{Node Classifications on the spatio-temporal dynamic graph $\mathcal{G}$}
\label{section_4}

\subsection{Brief introduction to the Spectral Graph Wavelet Transform}

Let $\mathcal G=(\mathcal V, \mathcal E, \mathcal W)$ be the spatio-temporal graph constructed in the above section. The graph signal $\mathbf{X}$ on $\cV$ with $\bX(\tau_{i,t})=\bx_i(t)$ defined on (\ref{graphsig}) represents the confirmed cases of city $i$ in the $t$-th week. Clearly, the cardinality of $|\mathcal V|=\hat N:=NT$. 
The {\it graph Laplacian} is defined by 
$\mathcal L = D - \mathcal W$, where the degree matrix
$D = (d_{ij})$ is a diagonal matrix with entries $d_{ii}=\sum_k w_{ik}$. The graph Laplacian $\mathcal L$ is not symmetric as our pandemic graph is directed. We construct a symmetric matrix by $L=(\mathcal L+\mathcal L^*)/2$, which is the averaging of the graph Laplacian and its transpose.
$L$ has $\hat N$ non-negative, real-valued eigenvalues $\Lambda=\{\lambda_1, \lambda_2, \cdots, \lambda_{\hat N}\}$ with $\lambda_1=0$, and the corresponding normalized eigenvectors $\{\bu_l, l=1,\cdots, \hat N\}$ form an orthonormal basis for the Hilbert space.
The graph Fourier transformation of $\bX$ can be written as
$$\hat\bX(\lambda_l)=\sum_{\tau\in \cV} \bX(\tau) \bu_l(\tau).$$
A {\it graph spectral filter}, or {\it kernel} $\hat{g}: \Lambda \rightarrow \mathbb{R}$ is a function defined in the spectral domain $\lambda_l \in \Lambda$.
A {\it dictionary} $\{\hat{g}_m\}_{m=1,2,\cdots, M}$ is a set of graph spectral filters constructed to detect different frequencies of signals.

In this paper, we use the dictionary constructed by Hammond et al.\cite{Hammond2011}:
\begin{equation}
\{\hat{h}(\lambda), \hat{g}(s_{M-1}\lambda), \hat{g}(s_{M-2}\lambda),\cdots, \hat{g}(s_1\lambda)\},\label{dictionary}
\end{equation}
where the wavelet kernel 
\begin{equation*}
\widehat{g}(\lambda)=\left\{
\begin{array}{ccc}
\lambda^2, & & 0 \leq \lambda<1,\\
-5+11\lambda-6\lambda^2+\lambda^3, & & 1\leq \lambda\leq 2,\\
4\lambda^{-2}, & & \lambda>2\\
\end{array} \right.
\end{equation*}
is a bandpass filter defined on the Fourier domain, the stretching scale $s_1, s_2, \cdots, s_{M-1}$ are logarithmically sampled between $s_1=1/\lambda_{\hat{N}}$ and $s_{M-1}=40/\lambda_{\hat{N}}$, the scaling function
$\widehat{h}(\lambda)=b \exp\left(-\left(\frac{10\lambda}{0.3\lambda_{\hat{N}}}\right)^4\right)$
is a low-pass filter detecting the low frequency signal with the parameter $b=\max_{\lambda}\hat{g}(\lambda)$. Such dictionary is designed to evenly cover the graph spectrum domain. 

Given a graph signal $\bX$ and the dictionary $\{\hat{g}_m\}_{m=1,2,\cdots, M}$, the spectral graph wavelet coefficient $W_{\bX}: \{1,\cdots,M\}\times \cV\rightarrow \mathbb{R}$ is defined as 
\begin{equation}\label{Wmtau}
W_{\bX}(m, \tau)=\sum\limits_{l=1}^{\hat{N}} \widehat{g}_m(\lambda_l)\widehat{\bX}(\lambda_l)\bu_l(\tau).
\end{equation}

However, the above SGWT is rather expensive for large graphs, as the resulting computational complexity is of order $O( {\hat N}^3)$. In this paper, we apply the fast spectral graph wavelet proposed in \cite{Hammond2011} to overcome this difficulty. The fast spectral graph wavelet transform based on Chebyshev polynomials approximation \cite{shuman2018distributed} shows that the graph wavelet coefficients in (\ref{Wmtau}) can be approximated by 
\begin{eqnarray}\label{Wbxm}
W_{\bX}(m, \tau) \approx \bigg(\frac{1}{2}c_{m,0} \bX+\sum\limits_{k=1}^{K_m} c_{m,k} \overline{T}_k(L) \bX \bigg)(\tau), 
\end{eqnarray}
where $K_m$ is the number of truncating terms, and
$$c_{m,k}=\frac{2}{\pi}\int_0^{\pi}\cos(k\theta)\hat{g}_m(\cos(\theta))\mathrm{d}\theta.$$ Moreover, $\overline{T}_k$ is the shifted Chebyshev polynomials with the domain of $[0,\lambda_{\hat{N}}]$, such that 
$\overline{T}_k(x)$ satisfies the recursive formula 
$$\overline{T}_k(x)=\left(\tfrac{4}{\lambda_{\hat{N}}} x-2\right)\overline{T}_{k-1}(x)-\overline{T}_{k-2}(x), $$ with initial conditions
$$\overline{T}_0(x)=1,\,\,\,\overline{T}_1(x)=\tfrac{2}{\lambda_{\hat{N}}}x-1.$$
The computational cost to approximate the wavelet coefficients is order $O(K_m |\mathcal E |+K_m |\mathcal V| )$.

In this paper, we take $K_m=40$ in (\ref{Wbxm}). It takes 0.5 seconds on a laptop with a 2.3 GHz 8-core Intel Core i9 processor. Without Chebyshev approximation, it costs 650 seconds.

\subsection{Node Classification Using SGWT coefficients}
Given our spatio-temporal dynamic graph $\mathcal G=(\mathcal V,\mathcal E,\mathcal W)$ together with the COVID-19 confirmed cases $\bX$, one question is how to provide a high-quality classification of nodes according to the pandemic spreading patterns? We refer to this as the ''node labeling problem'' with the understanding that the node classification problem can be abstracted as providing a classification label for the graph structure.
Here we will use the information already encoded in the SGWT coefficients to help us predict labels.

Let $T = 41$ be the number of time slices in the spatio-temporal dynamic model from December 6, 2020, to September 25, 2021. Let $M = 8$ be the number of graph spectral filters. Using the SGWT on the graph signal $\bX$, one gets a $\hat N\times M$ wavelet coefficient matrix with entires (\ref{Wbxm}). We then define node classification criteria based on the multiresolution information carried by SGWT coefficients. We will classify the nodes by the torque value obtained by some weighted average of the SGWT coefficients. 

\noindent {\bf Step 1: Data processing of wavelet coefficients.}

Firstly, we use a data scaling method to ensure for each wavelet coefficient that guarantees that they are on the same scale. This method was introduced by RobustScaler\cite{2017Introduction} in the machine learning library Scikit-learn as follows:
\begin{equation*}
S(m,\tau)=|W_{\bf{X}}(m,\tau)|/R(m),
\end{equation*}
where $R(m)$ is the interquartile range of $|W_{\bf{X}}(m,\tau)|$ with respect to $\tau$. The interquartile range is the difference between the first and third quartile of a data set. The quartiles make the RobustScaler ignore data points that are very different from the rest (like machine errors).

Secondly, we use logarithmic normalization to normalize each coefficient to [0, 1] to facilitate the calculation of the torque value and comparison,
\begin{equation*}
\overline{W}_{\bf{X}}(m,\tau) =\frac{\ln\left(1+S(m,\tau)\right)}{\ln\left(1+\max_{\tau}S(m,\tau)\right) }.
\end{equation*}
Logarithmic normalization does not change the order of the values, but it can significantly reduce the effect of maximum abnormal value \cite{yao2019developing}.

\noindent {\bf Step 2: Spatial node classifications based on torque values.}
In order to use the SGWT coefficients to classify the nodes, we need to define the torque value function $\varphi:\mathcal V\to\mathbb{R}$ by:
\begin{equation*}
\varphi(\tau) = [\overline{W}_{\bf{X}}(1, \tau),\cdots,\overline{W}_{\bf{X}}(8,\tau)]\cdot [-4,-3,-2,-1,\ 1,\ 2,\ 3,\ 4].
\end{equation*}
High torque values imply a high frequency of signal changes \cite{DalCol2018}. More precisely, if a vertex $\tau(i,t)$ has a relatively high torque value, then there are anomalies in the pandemic spread patterns of the $i$-th city compared to its surrounding cities at the time $t$.

Let $d =\max_{\tau}\varphi(\tau) -\min_{\tau}\varphi(\tau)$. 
By splitting the interval $[\varphi_{\min}, \varphi_{\max}]$ into 5 equal intervals, the vertex space $\cV$ can be divided into five disjoint sets $\{\cV_1, \cV_2, \cV_3, \cV_4, \cV_5\}$, which represent the spatio-temporal changes of the $\bx_i(t)$ (in terms of the SGWT coefficients) as low frequency, mid-low frequency, uncertain, mid-high frequency, and high frequency, respectively.
The detailed node classifications are defined in Table \ref{tab:node_class}.
 
\begin{table}[htbp]
\centering
\caption{Node classification in $\cV$. }
		\label{tab:node_class}
\resizebox{1.0 \columnwidth}{!}{ 
\begin{tabular}{|c|c|c|c|c|c|}
\hline
Torque Value Range & [$\varphi_{\min}$, d/5] & (d/5, 2d/5] & (2d/5, 3d/5] & (3d/5, 4d/5] & (4d/5, $\varphi_{\max}$] \\ \hline
Node Subset & $ \cV_1$ & $ \cV_2$ & $ \cV_3$ & $ \cV_4$ & $ \cV_5$ \\ \hline
Signal Frequency & low & mid-low & uncertainty & mid-high & high \\ \hline
\end{tabular}
}
\end{table}

\begin{figure}[htbp]
     \centering
	\includegraphics[width=9cm,height=5.5cm]{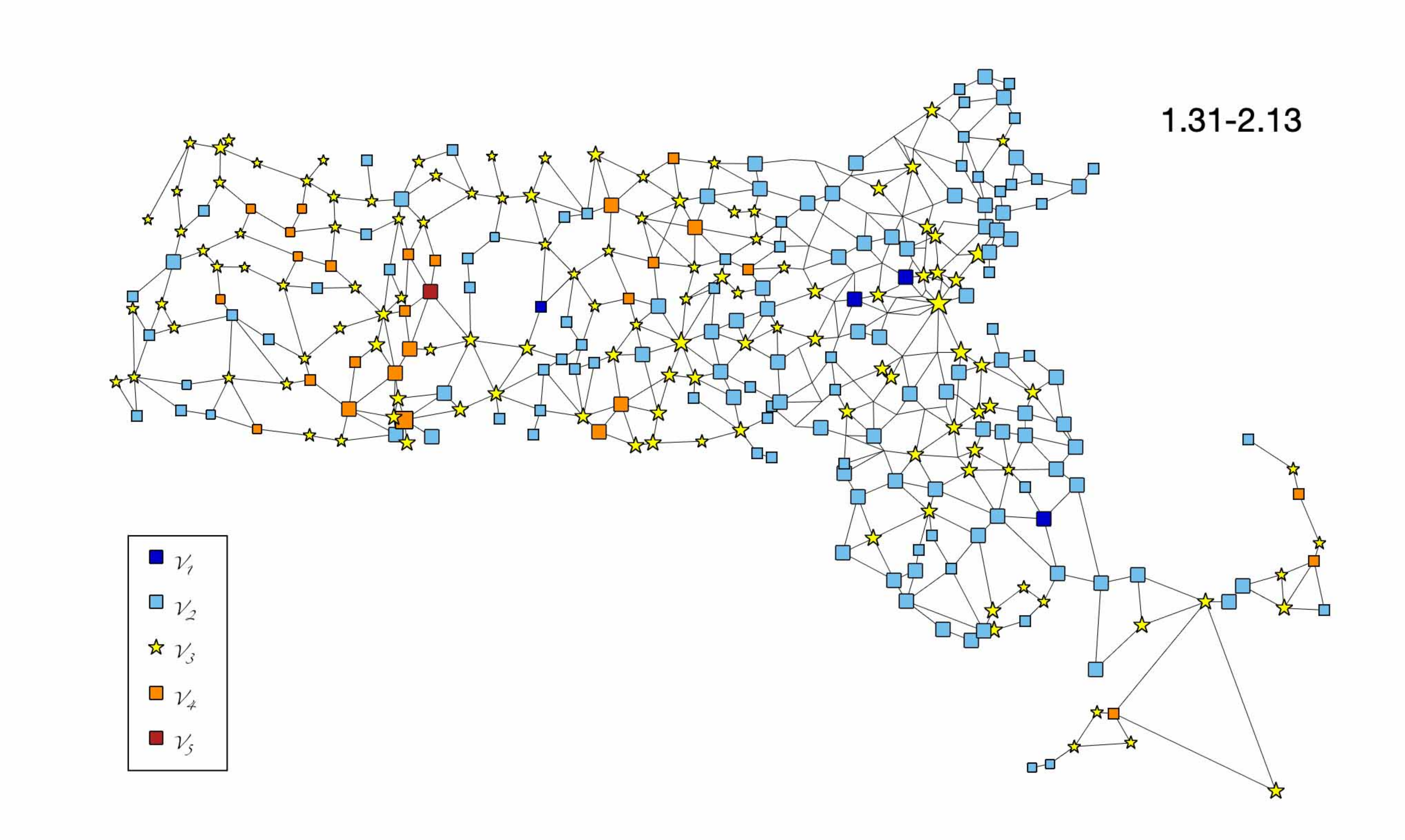}
     \caption{Nodes classification of the 9th time slice. The dark red square represents Amherst.} \label{class}
\end{figure}

The spatial node classification method is demonstrated in Fig. \ref{class}, for the 2-week period of Jan.31-Feb. 13, 2021.

\begin{figure}[htbp]
\centering
\includegraphics[width=9cm]{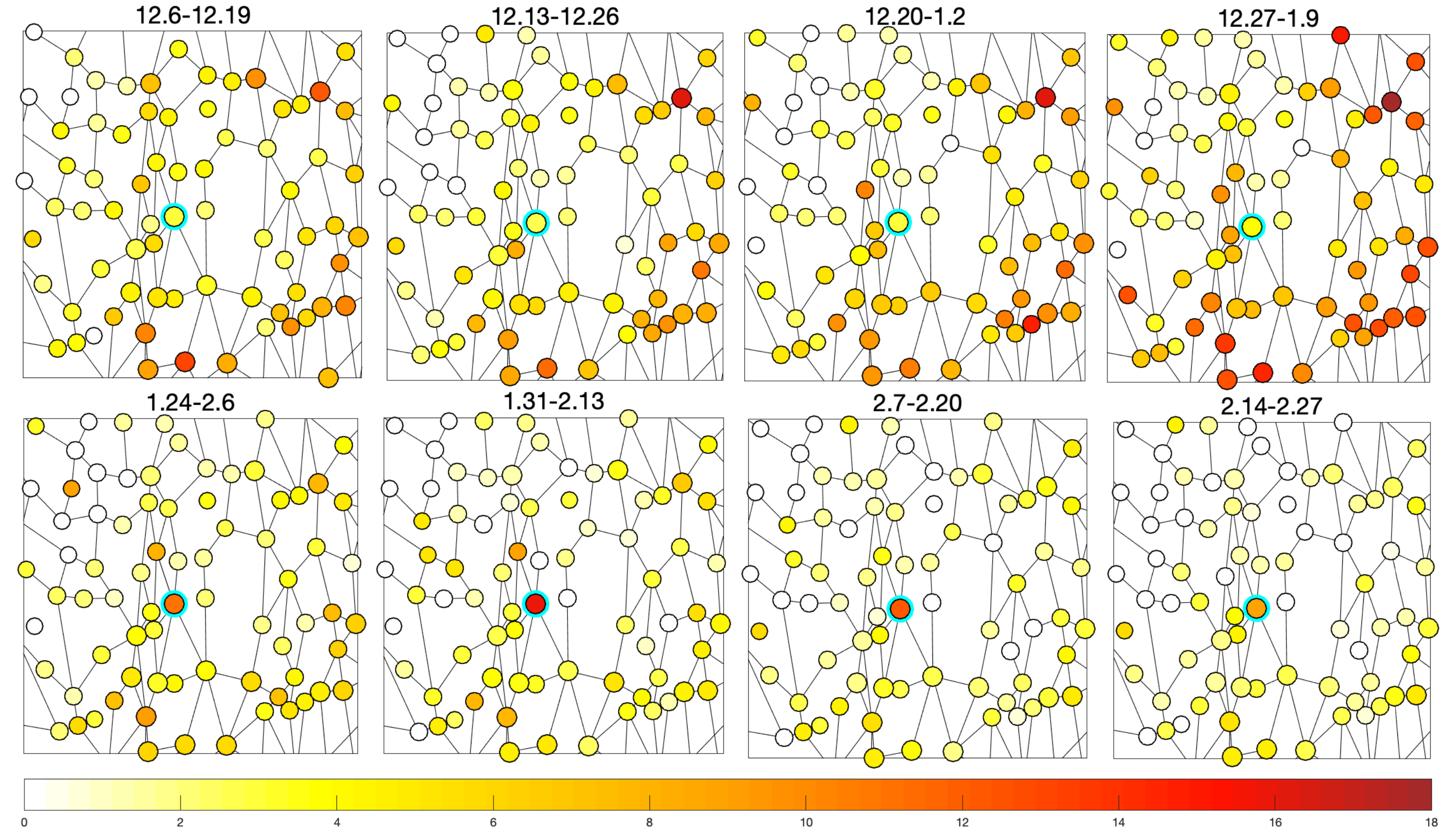}
\caption{Visualization of the pandemic graph signal $\bX$ at Amherst and its adjacent areas from Dec. 6, 2020, to Jan. 9, 2021 and Jan. 24, 2021, to Feb. 27, 2021. The color represents the actual value of the signal. The nodes marked by the blue circle represent the city of Amherst.}
\label{fig_node8_signal}
\end{figure}

\begin{figure*}[htp]
\centering
\includegraphics[width=17cm,height=6.5cm]{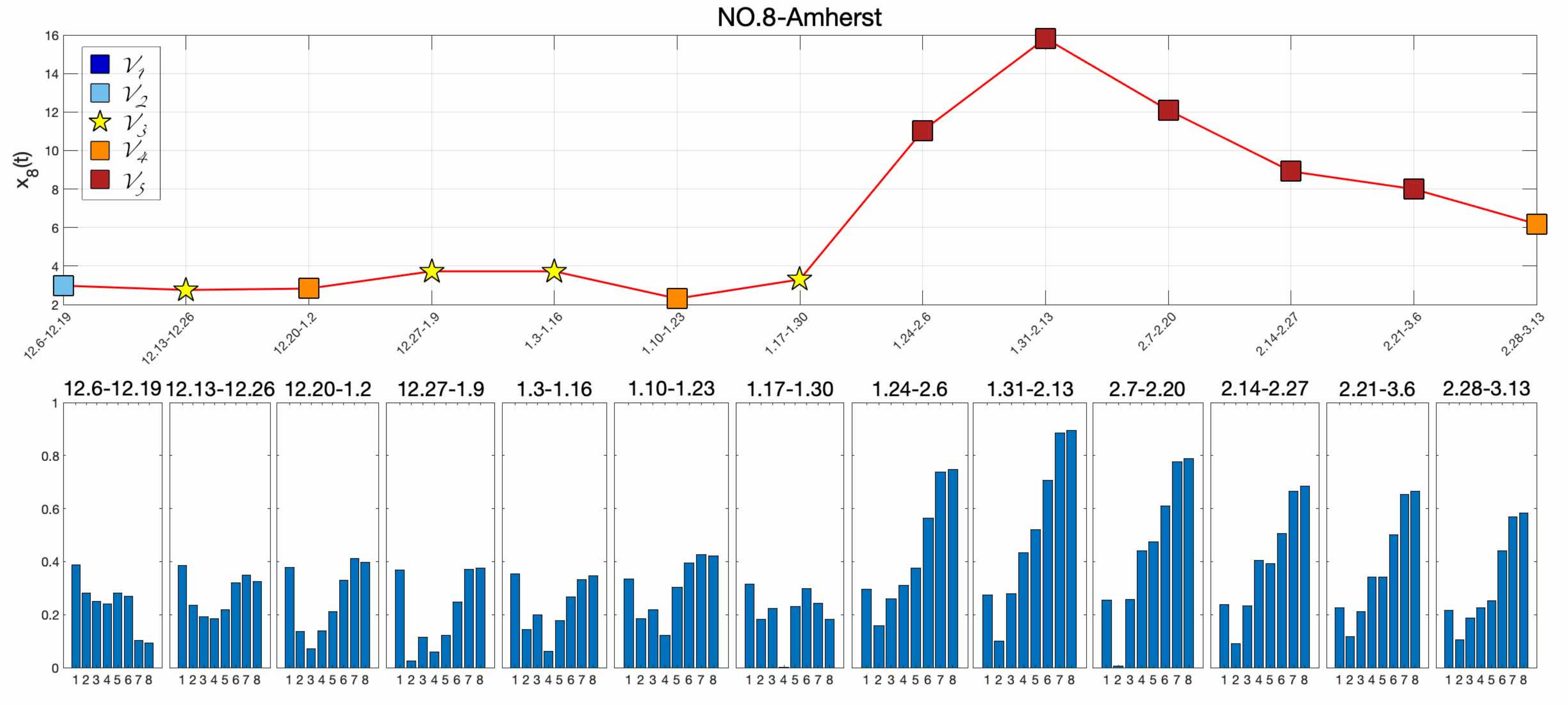}
\caption{ Nodes classification (top) and normalized graph wavelet coefficients (bottom).}
\label{node_bar}
\end{figure*}

An efficient node classification should be able to capture the abrupt changes of the graph signal, both spatially and temporally. To make the illustrations more explicit, we use our node classification method to investigate the city of Amherst during the 13 weeks from December 16, 2020, to March 13, 2021. Fig. \ref{fig_node8_signal} shows that the confirmed cases in Amherst were significantly higher than its neighbors from January 24, 2021 to February 27, 2021. Fig. \ref{node_bar} demonstrates the different temporal stages of the node classification value of Amherst during the 13 weeks.
Compared with Fig. \ref{fig_node8_signal}, we can see that when the City of Amherst has similar pandemic patterns as the surrounding cities, the wavelet coefficient represents low frequency (Dec. 6-Dec. 19). On the other hand, as the confirmed cases in the surrounding cities start to rise, the wavelet coefficient of Amherst represents high frequency (Dec. 20-Jan. 2), which means that the pandemic cases of Amherst have different patterns than its neighbors. Our node classification method accurately captures the abnormal pandemic spread patterns of Amherst during the period from January 24, 2021, to March 6, 2021, see Fig. \ref{fig_node8_signal} and Fig. \ref{node_bar}.
In this period, the University of Massachusetts Amherst partially opened its campus, and many students returned to Amherst \cite{umass1}.

This method can effectively identify the pandemic spread anomaly patterns of any specific period, which should be valuable for further research. 


\section{Visualization Analysis of COVID-19 Pandemic Spread Patterns}
\label{section_5}

In this section, we focus on the visualization analysis of the COVID-19 pandemic spread patterns on the spatio-temporal dynamic graph model.
We first conduct an overall analysis of the COVID-19 data in Massachusetts. 
Then we identify the supper pandemic spreaders (cities). Finally, a refined node classification is performed in $\mathcal V_4$ and $\mathcal V_5$, where the pandemic spread patterns have abrupt spatio-temporal changes. 

 \subsection{Overall graph classifications}
To have an overall temporal analysis of the pandemic evolution patterns for cities in Massachusetts, we define a graph classification method for the spatio-temporal dynamic graph $\cG$, which enables us to classify each graph time slice for $\{G_t, t=1,\cdots, T\}$. 

First, we need to calculate the node classification distribution in $G_t$, which can be obtained by counting the frequencies of nodes in $G_t$ that occur in each of the five classes $\cV_j$, $j=1,\cdots, 5$: 
\begin{equation}\label{distc}
\sigma_t^{j}=\frac{1}{N} \sum\limits_{i=1}^{N} \mathbb{I}_{ \mathcal V_j}(\tau_{i,t}), j=1,2, \cdots, 5,
\end{equation}
where $\mathbb{I}_{ \mathcal V_j}$ is the indicator function of the set $\cV_j$. The distribution $\{\sigma_t^j, j=1,\cdots, 5\}$ can not be used to directly classify the graph $G_t$, as one has to weigh in the distributions in other time slices to well capture the role of each frequency class. We adopt the method used by Dal Col. etc.\cite{DalCol2018}, by first taking $\sigma_{\max}^{j} = \max \{ \sigma_1^{j}, \sigma_2^{j}, \cdots, \sigma_T^{j} \}$. Then we define the class of $G_t$ as $ \mathcal V_{r_t}$, with 
\begin{equation}\label{timeslice_class}
	r_t=\mathop{\arg\max}\limits_{j\in\{1,2,3,4,5\}}( \sigma_t^{j}/\sigma_{\max}^{j}).
\end{equation}

\begin{figure*}[htp]
\includegraphics[width=17cm,height=8.5cm]{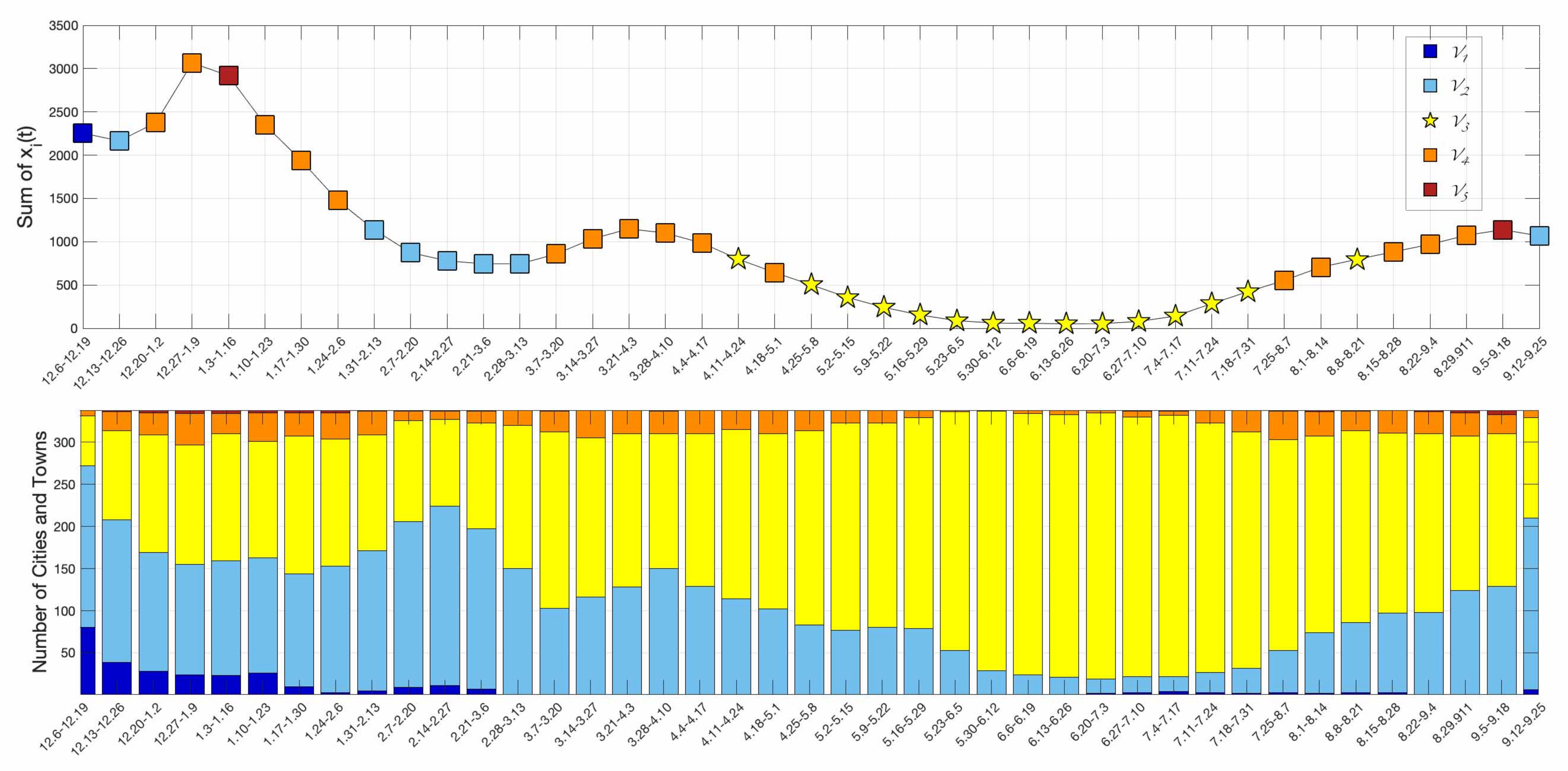}
\caption{Overall visual analysis of COVID-19 in Massachusetts. The figures depict graph classification for each graph time slice ({\bf top}) using method (\ref{timeslice_class}), and the spatial node classification distribution $\sigma_i^j$ given by (\ref{distc}) ({\bf bottom}).}
\label{network_analysis}
\end{figure*}

Next we conduct a temporal graph classification to get an overall visual analysis of the spatio-temporal pandemic spread patterns in Massachusetts, which is demonstrated in Fig. \ref{network_analysis}.
As we can see from Fig. \ref {network_analysis}, after Thanksgiving, 2021, Massachusetts saw a substantial spike in confirmed cases as there were more group gatherings. The confirmed cases in Massachusetts reached their peak during the Christmas holiday period. We can infer that multi-site outbreak patterns exist because the temporal and spatial changes of confirmed cases in many cities are different from those in the surrounding cities.
The cases declined sharply after mid-January due to restricting gatherings, maintaining social distance, and advocating masks, then stabilized in late February (mid-low frequency). Massachusetts entered the fourth phase of the state's reopening plan on March 22, 2021.
With the relaxation of pandemic restrictions, there was a small-scale rebound of confirmed cases with the multi-point pattern at the end of March (mid-high frequency). The numbers began to decrease from April to July due to the increased vaccination rate. 

\subsection{Spatial ranking for super-spreader cities}
Identifying influential spreaders (cities) in a pandemic network is a core question to prevent pandemic spreading. Many different centrality measures tied to the network topology have been introduced to find the central nodes. However, the importance of node features is mostly ignored in these works. The centrality of a node in a pandemic network should depend on two features: its local influence on the nodes in its one-hop neighbor and its global influence on the nodes belonging to higher-order hop neighbors. 
Taking advantage of the attention coefficients learned from the GAT, we introduce a new score function that allows us to redefine the classical centrality measure by considering their importance to the neighbors' features.

Considering the $m$-hop neighbor of node $\tau_i$ in the MAR graph $G_{MAR}=(V,\widetilde{E},P)$.
The attention coefficient measures the importance of $\tau_i$'s features to $\tau_j$. Let $P^m=({p}_{i,j}^{(m)})_{N\times N}$ be the $m$-th power of the matrix $P$. Then ${p}_{i,j}^{(m)}$ characterizes the m-step influence of city $i$ to city $j$. We define a new centrality metric to capture the capability of city $i$ to spread the pandemic to their neighbors by
\begin{equation*}
c_i^{(m)}=\sum_{j}{{p}}^{(m)}_{ji}, j\neq i, m=1, 2, 3, \cdots.
\end{equation*}
In our practice, we define the {\it{influential score}} of city $i$ by
$$C_i=\sum_{m=1}^5 c_i^{(m)}$$
as the total capability of city $i$ to spread the virus to other cities. 
 
\begin{figure*}[htbp]
\centering
	\includegraphics[width=15cm,height=8cm]{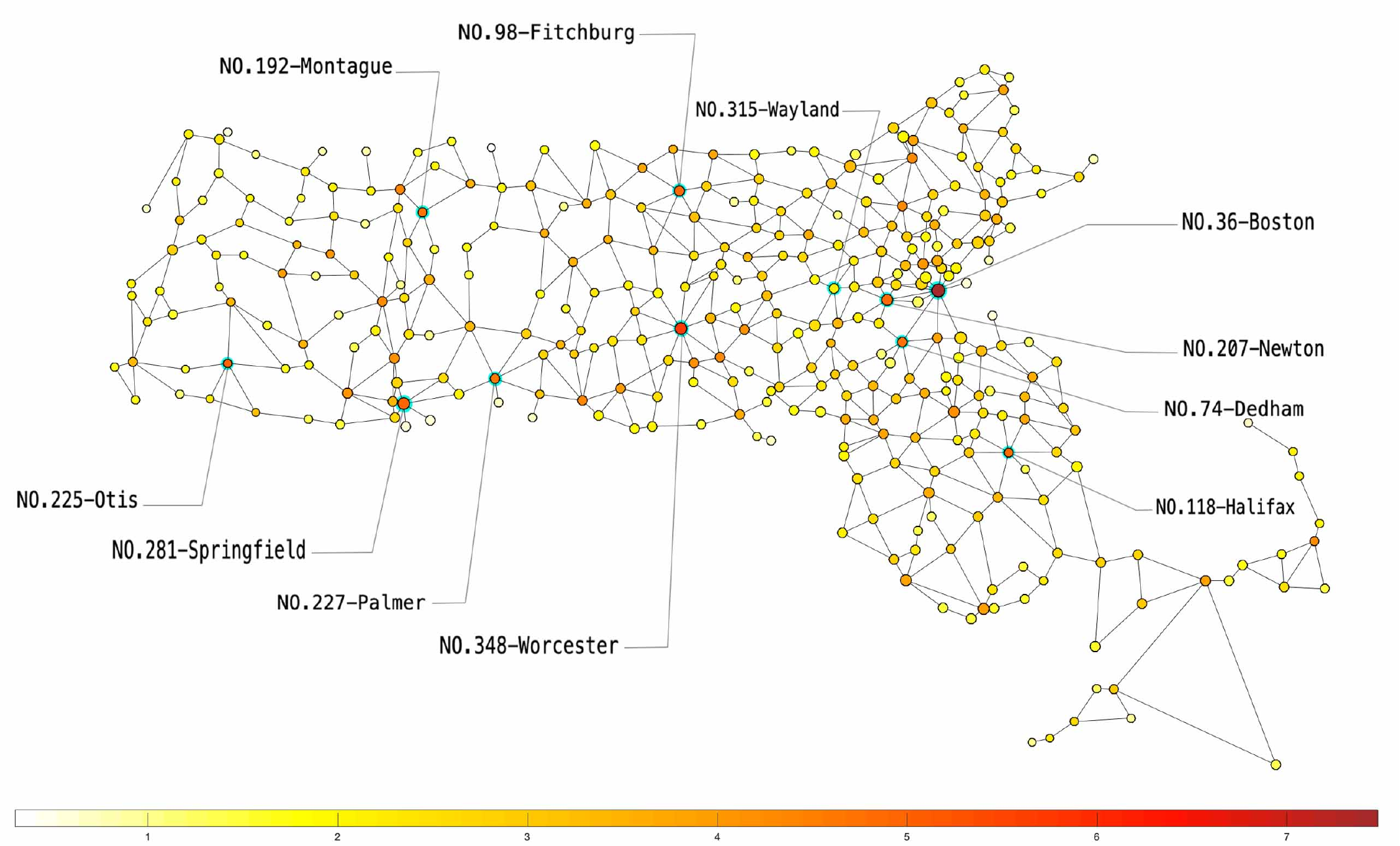}
  \caption{City spatial ranking of pandemic spread capability: from dark red (top rank) to white (low rank).} \label{influ_rank_local}
\end{figure*}

The spatial ranking of cities according to their influential scores is shown in Fig. \ref{influ_rank_local}. One can see that the top ten cities with the largest pandemic spread (influential) scores (in dark red) are Boston, Worcester, Springfield, Otis, Fitchburg, Newton, Halifax, Dedham, Montague and Palmer. 
It's not surprising that Boston, Worcester, and Springfield are in the top three, given these cities' large populations and high mobility. On the other hand, we are able to identify small cities, which are otherwise difficult to identify without using our advanced methods. For example, the City of Otis has a small population but ranked fourth on the list. It is remarkable that the City of Wayland has a larger population and the same number of connected US Routes compared to Otis, but it ranks much lower. This has been confirmed by news in \cite{news20210917}, as well as the fact that Otis has a ski resort. Using this new ranking metric, we can correctly identify cities at risk and those more volatile to pandemics and thus need to invest more social resources to better control the pandemic.

\subsection{Refined node classification in anomaly cities}

Given the COVID-19 data in Massachusetts, another important question is to identify cities with anomaly spread patterns during the pandemic period. More precisely, this is equivalent to finding patterns in data (i.e., attribute values or changes in the values over time) that are significantly different from that of the spatio-temporal neighbors.
These non-conforming patterns are often referred to as anomalies. The typical anomaly detection method concentrates on time series, or static graph signals, which could not be applied to our case. We aim to design a new method that enables us to detect anomalies in COVID-19 time-series data belonging to multiple entities (cities) and explain the anomalies to domain experts in a comprehensible manner.

\begin{figure}[htbp]
\includegraphics[width=9cm,height=4.5cm]{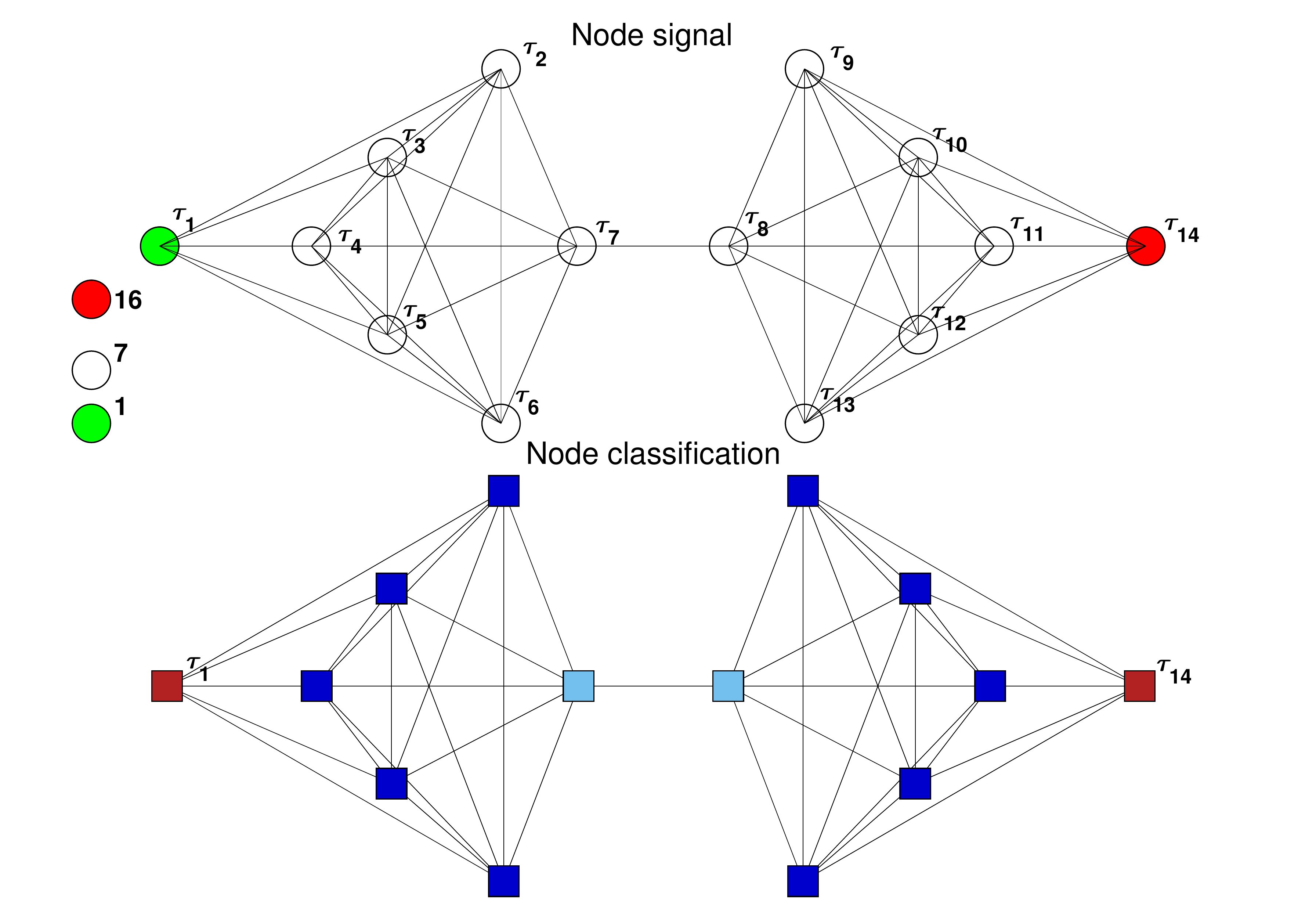}
\caption{A specially-designed graph signal function $f: V\to \{1,7,16\}$, such that $f(\tau_1)=1$ and $f(\tau_{14})=16$. Both $\tau_1$ and $\tau_{14}$ are classified in class $\cV_5$, even though they are completely different typies of outliers.
}\label{fig_class_compare}
\end{figure}

SGWT is a powerful tool for finding anomalies for spatio-temporal graph signals. For example, the set $\mathcal V_5$ can be counted as one anomaly class, both spatially and temporally. 
However, in $\mathcal V_5$, by examining the distribution of the SGWT coefficients more carefully, we find that there are situations in which there is a node with a higher or lower signal value than that of the surroundings. 
As shown in Fig. \ref{fig_class_compare}, there are two dramatically types of vertices in $\mathcal V_5$, as indicated above. 
 
Below we say the nodes in $\cV_4\cup \cV_5$ are anomaly nodes. The corresponding city is the anomaly city. We use a refined node classification method to identify these two abnormal situations in anomaly cities.
Let $\mathcal{N}_i^{(1)}=\{\tau_j\in V\,|\, a_{i,j}=1, j \neq i\}.$
 We first define an anomaly metric:

\begin{equation}\label{mit}
\vartheta (i,t)=\left\{
\begin{aligned}
\frac{\bx_i(t) }{\frac{1}{|\mathcal{N}^{(1)}_i|}\sum_{j\in {\mathcal{N}}^{(1)}_i } \bx_j(t)}, &\quad \sum_{j\in {\mathcal{N}}^{(1)}_i } \bx_j(t) \neq 0\\
\max\{\bx_i(t),1\}, &\quad \sum_{j\in {\mathcal{N}}^{(1)}_i } \bx_j(t) =0,
\end{aligned}
\right.
\end{equation}
which is the relative ratio of the data information of city $i$ at time $t$ compared with the averaging data obtained from its one-hop neighbors. 

\begin{figure}[htbp]
\includegraphics[width=8.5cm]{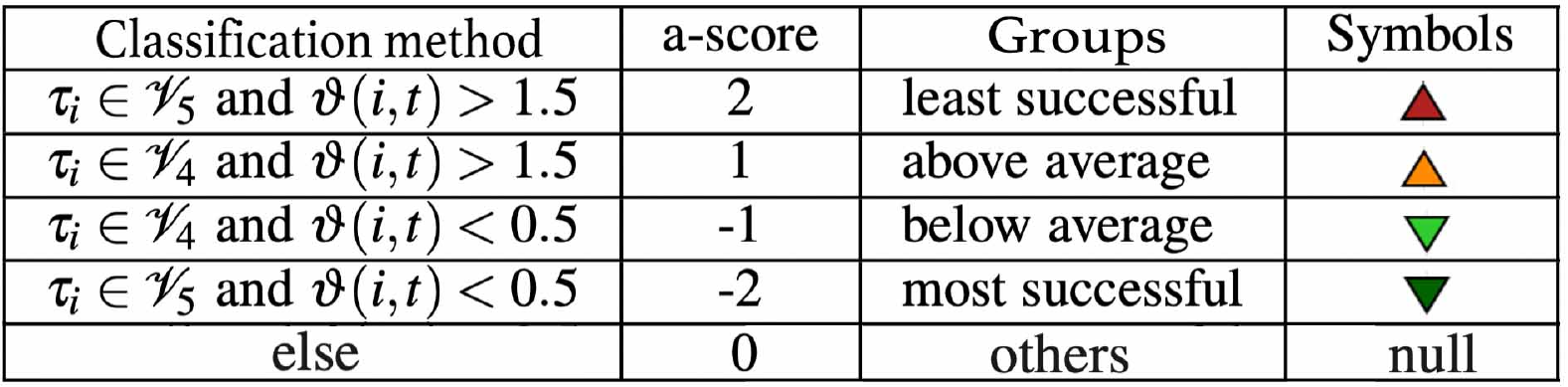}
   \caption{Refined node classification and a-score in $\mathcal V_4 \cup \mathcal V_5$. } \label{tab:city_score}
\end{figure}

Then we define the anomaly score (a-score) $a(i,t)$ to the nodes in the set $\mathcal V_4 \cup \mathcal V_5$ according to Fig. \ref{tab:city_score}. According to the a-score, we divide the $\cV_4\cup \cV_5$ into five groups as in Fig. \ref{tab:city_score}.
 
A city with high a-scores presents at risk or volatile in the pandemic. It should ring an alarm for local city administrations, as the city has the potential to increase infection spread to its neighboring cities. Earlier identifying these cities can prevent them from becoming the outbreak cities of a pandemic.
On the other hand, identifying cities with low a-scores can be valuable in controlling virus spread, as they provide good models for other cities.

\begin{figure}[htbp]
\includegraphics[width=9cm]{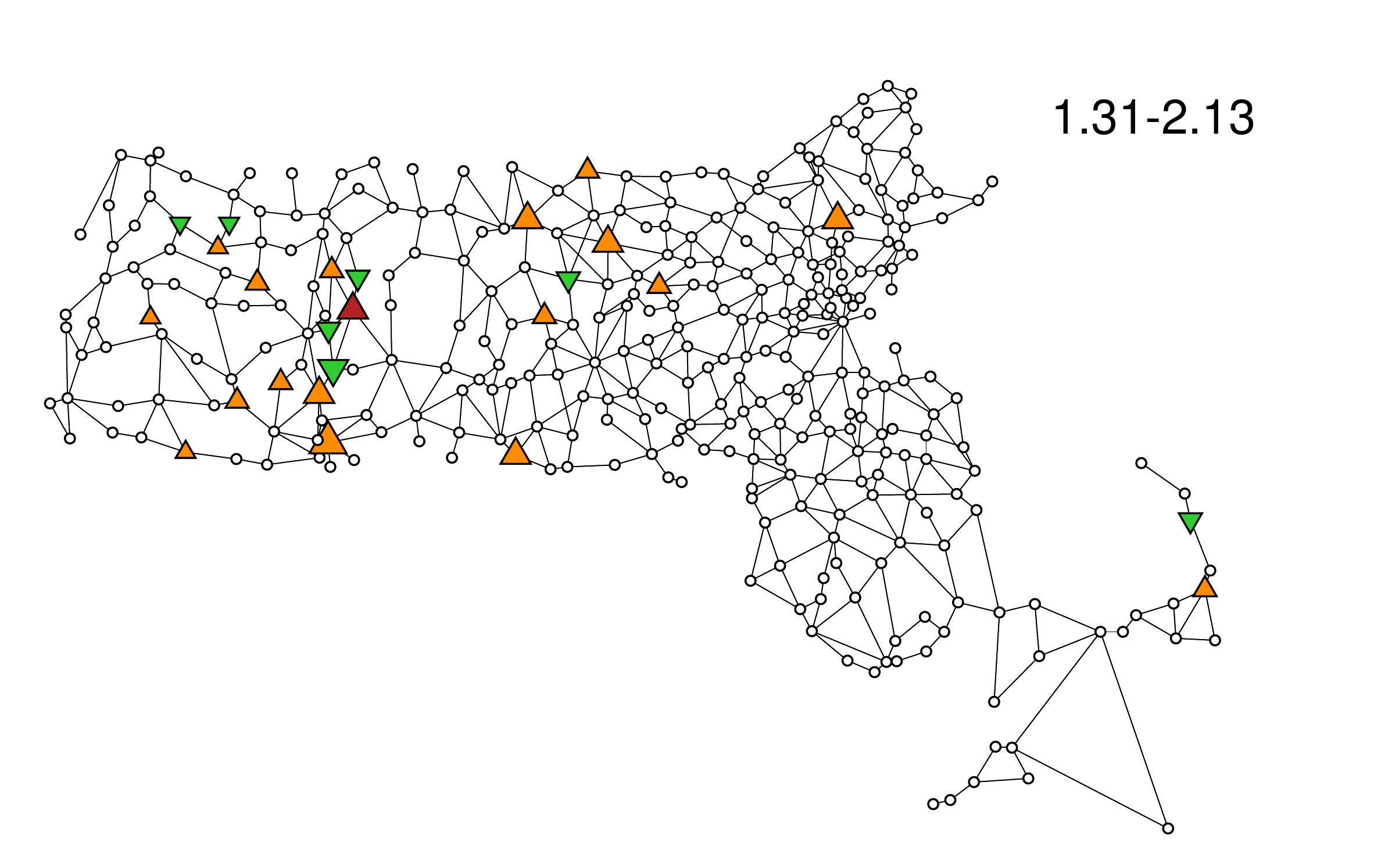}
   \caption{Refined node classification for the 9th week, from Jan. 31 - Feb. 13.
}\label{newclass}
\end{figure}
Fig. \ref{newclass} depicts the data visualization plots for cities with anomaly patterns from January 31 to February 13. We can quickly detect Amherst with a high a-score, which is also confirmed as one of the most significant COVID-outbreak cities in Fig. \ref{node_bar}.

\subsection{Spatio-temporal ranking of super-spreader cities}

The a-score defined in Fig. \ref{tab:city_score} depends on both spatial and temporal coordinates. To have an overall picture of the pandemic evolution behavior of these cities and evaluate the urban pandemic prevention and control, we propose a data visualization plot to explain the level of anomalies by the anomaly score. 
Below, we use a-scores in Fig. \ref{tab:city_score}, to provide a new city ranking, by taking averages of its a-scores over the 41-week period for each city:
\begin{equation}\label{a_score}
\bar a(i):=\frac{1}{41}\sum_{t=1}^{41} a(i,t).
\end{equation}
\begin{figure}[htbp] 
      \centering
	\includegraphics[width=9cm]{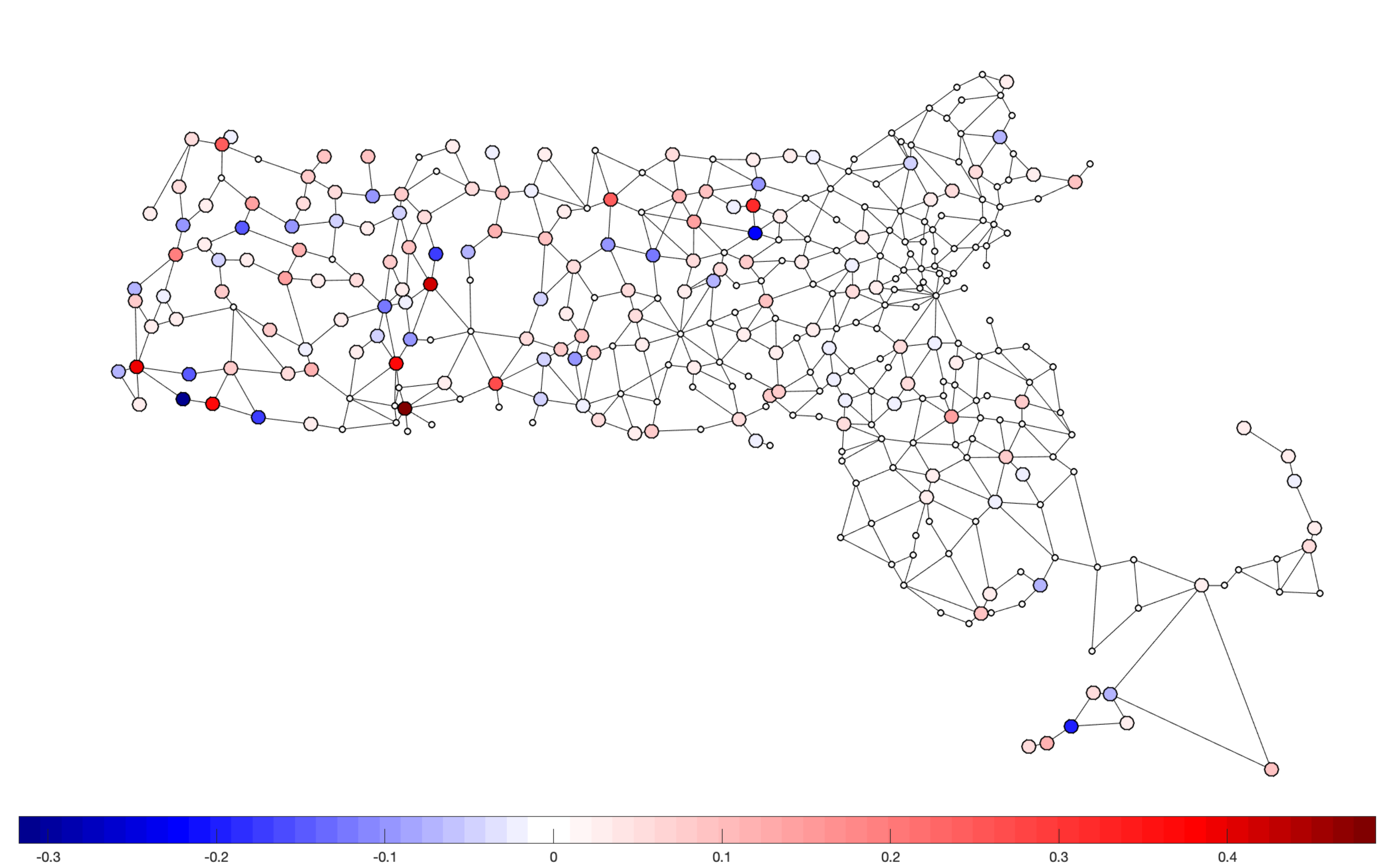}
\caption{City spatio-temporal ranking from Dec. 6, 2020 to Sept. 25, 2021, using the average a-score $\bar a$. Dark red refers to the highest average a-score $\bar a$, while dark blue refers to the lowest average.}
\label{rankmap}
\end{figure}
Fig. \ref{rankmap} shows the average anomaly scores of each city in MA.
One can see that cities with higher $\bar a(i)$ values are more likely to become super-spreader (cities) for any future pandemic.
A by-product of using the $\bar a(i)$ value is that, one could evaluate the city's pandemic prevention and control capabilities. In Fig. \ref{rankmap}, the darker the red, the cities have the less successful pandemic prevention and control. On the other hand, the darker the blue, the cities are more successful in pandemic prevention and control. The top five least successful cities are identified as Springfield, Amherst, Great Barrington, Holyoke, and Sandisfield; as well as the top five most successful cities are New Marlborough, Harvard, West Tisbury, Tolland and Leverett.

\begin{figure*}[h]
\includegraphics[width=16cm]{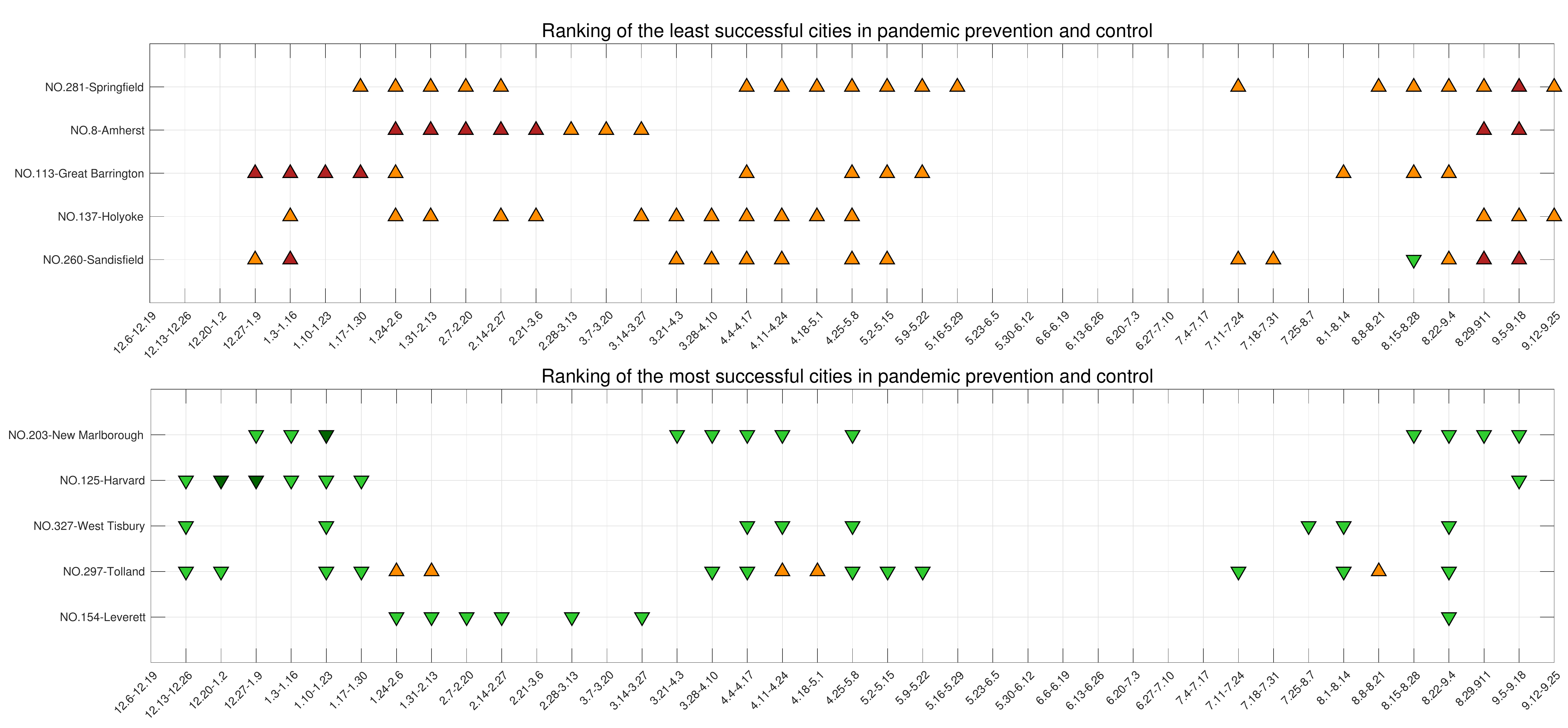} 
\caption{Ranking of the least successful and most successful cities in pandemic prevention and control from Dec. 16, 2020 to Sept. 25, 2021. } \label{noderank}
\end{figure*}

\begin{figure}[htbp]
\centering
\includegraphics[width=9cm]{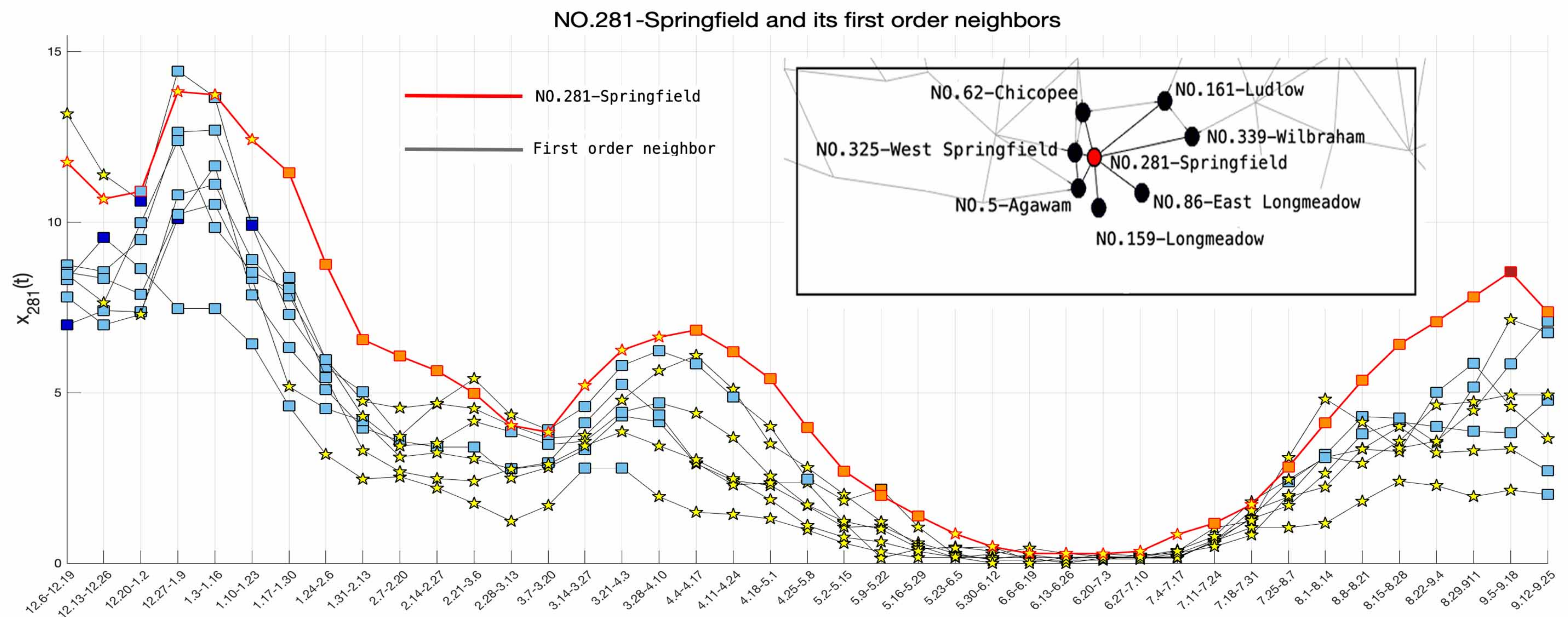}
\caption{The 41-week evolution of COVID-19 confirmed cases for Springfield and its neighboring cities from Dec. 19, 2020 to Sept. 25, 2021.}
\label{compare_high1}
\end{figure}

\begin{figure}[htbp]
\centering
\includegraphics[width=9cm]{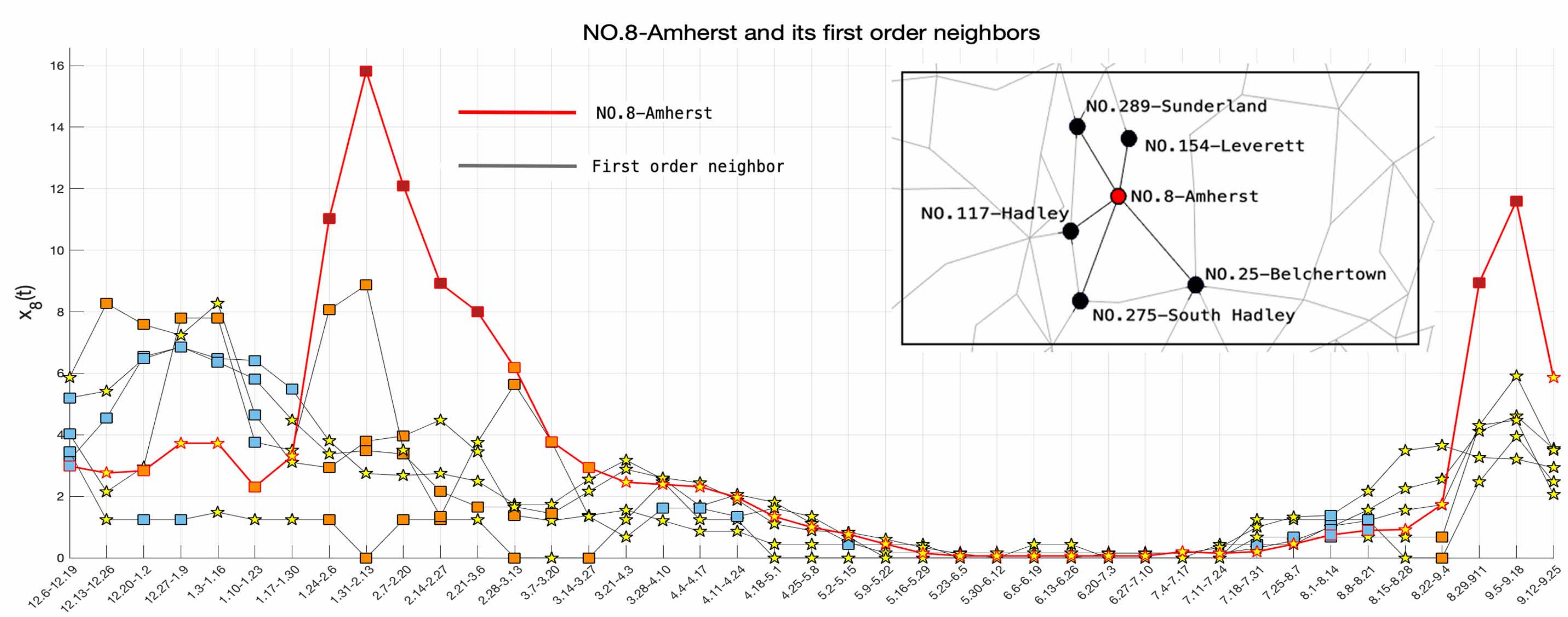}
\caption{The 41-week time evolution of COVID-19 confirmed cases for Amherst and its neighboring cities from Dec. 19, 2020 to Sept. 25, 2021.} \label{compare_high2}
\end{figure}

\begin{figure}[htbp]
\centering
\includegraphics[width=9cm]{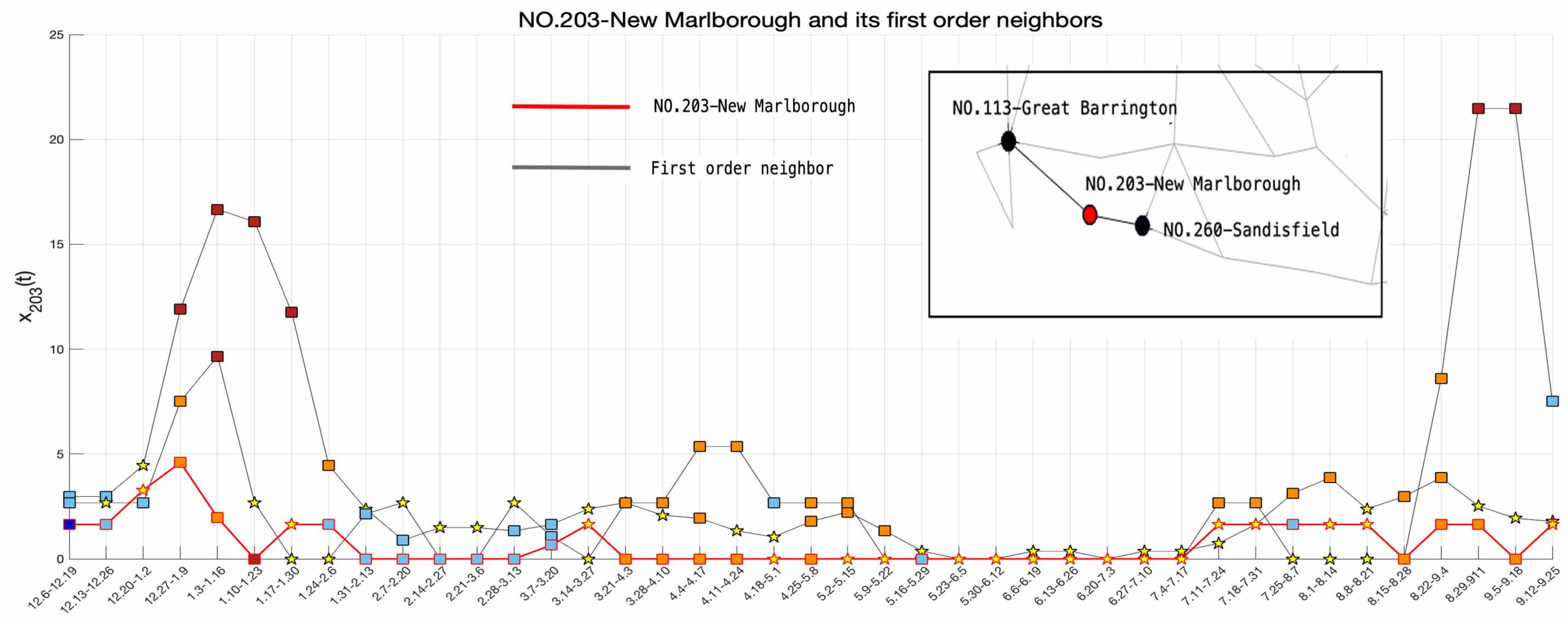}
\caption{The 41-week evolution of COVID-19 confirmed cases for New Marlborough and its neighboring cities from Dec. 19, 2020 to Sept. 25, 2021.} \label{compare_low1}
\end{figure}

\begin{figure}[htbp]
\centering
\includegraphics[width=9cm]{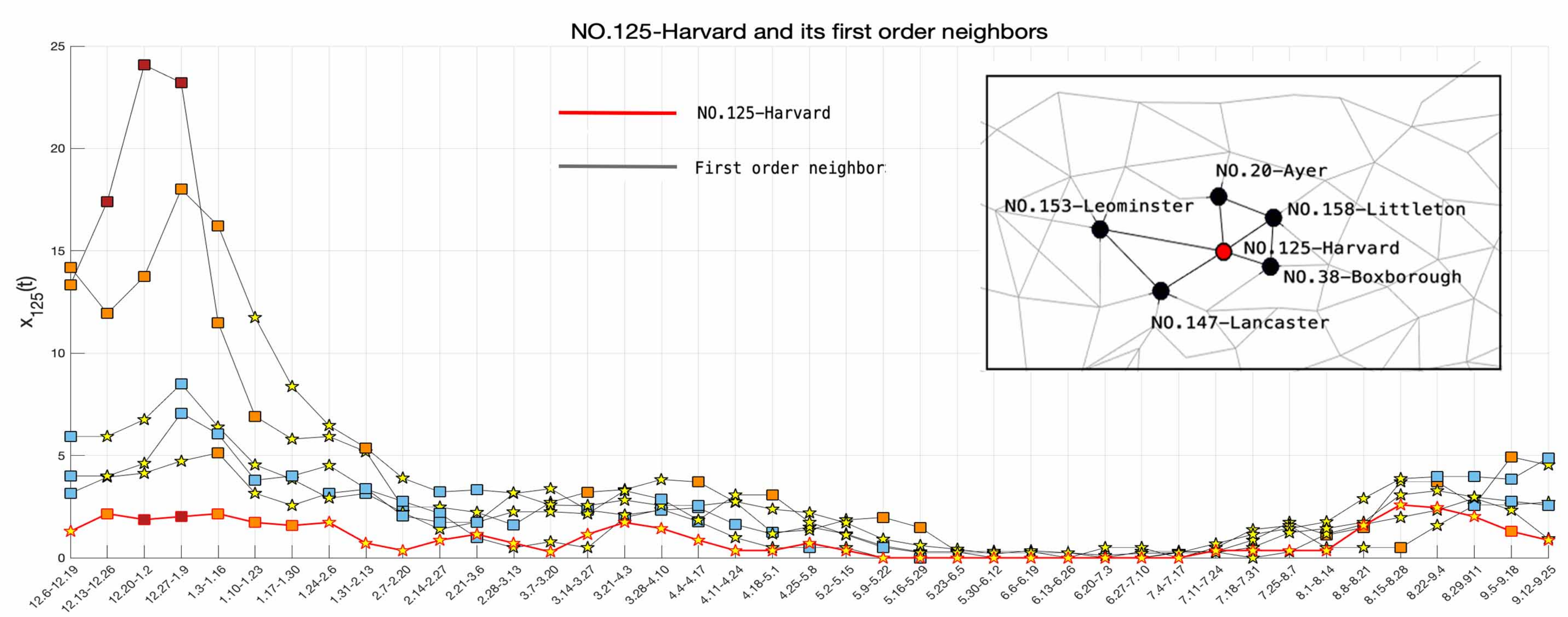}
\caption{The 41-week evolution of COVID-19 confirmed cases for Harvard and its neighboring cities from Dec. 19, 2020 to Sept. 25, 2021.} \label{compare_low2}
\end{figure}

The detailed visualization analysis for these cities is included in Fig. \ref{noderank}. This visual view can quickly locate the outbreak time and duration of abnormal phenomenons. To make effective comparisons, we also plot the pandemic time evolution data of Springfield, Amherst, New Marlborough and Harvard, see Fig. \ref{compare_high1} - Fig. \ref{compare_low2}.
 
Our analysis identifies Springfield as the least successful city using data from December 6, 2020, to September 25, 2021. Fig. \ref{noderank} shows that most of the confirmed cases in Springfield surpass the values of the surrounding areas during the entire 41-week period, which can also be confirmed in Fig. \ref{compare_high1}. A recent report shows that Springfield is a center for the Delta outbreak \cite{Springfield1}. This confirms that the city has a much higher risk of spreading new pandemic variants to other cities and may need more help from administrations to control the pandemic.

Amherst is ranked as the second least successful city for 41 weeks from December 6, 2020 - September 25, 2021. Fig. \ref{noderank} indicates that Amherst has much higher confirmed cases than neighbors from January 24 to March 27, 2021, which is consistent with Fig. \ref{compare_high2}.
Indeed, we show that Amherst ranks as the first pandemic spreader during a two-week short periodic from January 31 - February 13, 2021, see Fig. \ref{class}.
By digging into the news of this period, we find that the University of Massachusetts Amherst started the spring semester on February 1. Because many students moved into dorms or apartments in Amherst, there had been more than 398 active cases during that period. Contact tracing data record \cite{umass2} has shown that some students failed to follow social distancing and mask protocols in social and residential settings, promoting pandemic virus transmission. Subsequently, the university further strengthened its management standards. For example, students who do not wear masks will not be allowed on campus, and large group gatherings were not encouraged. The pandemic spread was effectively brought under control at the end of February 2021. At the beginning of the fall semester (the end of August 2021), a similar abnormal situation appeared again in Amherst, see \cite{amherstma1,umass3}.

New Marlborough and Harvard are the top two most successful cities for the 41 weeks from December 6, 2020, to September 25, 2021.
Fig. \ref{noderank} illustrates that the two cities have much lower confirmed cases than those in the surrounding cities most of the time. 
Fig. \ref{compare_low1} and Fig. \ref{compare_low2} describe the time evolution of the top two most successful cities, which have good and stable spread patterns during the pandemic, and are consistent with the situation illustrated in Fig. \ref{noderank}.
It indicates that these cities might have adequate prevention and control measures.
By digging into the news of this period, the Health Board in New Marlborough looked for a temporary part-time COVID Ambassador to provide COVID-19 prevention guidance to citizens, businesses and community groups \cite{marlborough2}, and held several sessions for discussions and updates on COVID-19 \cite{marlborough3}. Furthermore, New Marlborough has provided free masks to residents since August 20, 2020 \cite{marlborough1}.
These measures allow New Marlborough to remain unaffected even when its neighbors sharply all had a sharp increase in cases, see Fig. \ref{compare_low1}.
Harvard offered free safety training courses for workers \cite{Harvard}. Citizens volunteered to help the Board of Health open a local vaccine clinic \cite{Harvard3}. These measures may be the reason for the success of Harvard's pandemic prevention.

In Fig. \ref{noderank}, we show the five most successful and least successful cities. 
By changing the time range of the formula (\ref{a_score}), we can also find successful and least successful cities for any period.
As a by-product, our visualization analysis is based on the refined node classification for our spatio-temporal dynamical graph, so it captures correlations of pandemic evolution patterns among these cities.
Take Springfield as an example, see Fig. \ref{noderank}. Because the spatio-temporal outbreak patterns of Springfield and Holyoke are similar, we can infer a significant correlation between them. Indeed, we know that they both have large traffic flows and are close to each other, see Fig. \ref{map}, thus the patterns of their pandemic outbreaks should be highly related. Our method can quickly help experts locate cities with similar traits and apply insights from a city to inform prevention and control strategies in similar areas.

\section{Conclusion}
\label{section_6}

This paper draws a Massachusetts Route graph based on the latitude and longitude of 351 cities and towns and the main traffic routes passing nearby.
The pandemic transition probability is learned through semi-supervised deep learning, based on the Graph Attention Neural Network.
We construct a spatio-temporal dynamic model by strong-product for the time-varying data on the graph, which can better capture the spatio-temporal pandemic evolution patterns. 
With the help of SGWT, the COVID-19 pandemic spread patterns in Massachusetts from December 6, 2020, to September 25, 2021, are analyzed and visualized. 
In addition to the overall analysis of the temporal and spatial evolution of the pandemic data, we also identify cities with strong pandemic spread influence, such as the town of Otis. We also construct a new anomaly indicator to classify cities with abnormal pandemic situations, indicating cities at risk of virus transmission, and identifying cities that are likely to be infected in the next step. The five cities identified with the least successful pandemic prevention are Springfield, Amherst, Great Barrington, Holyoke, and Sandisfield; the five cities with the most successful pandemic prevention are New Marlborough, Harvard, West Tisbury, Tolland, and Leverett.
These obtained results not only offer descriptive insight for strategizing purposes in combating the COVID-19 outbreak in Massachusetts, but also can be used to protect those high-risk cities for the next possible round of pandemics. Further, it also provides a framework for evaluating regional pandemic prevention work in the COVID-19 pandemic. It helps the policymakers identify cities with better prevention in the pandemic and extract insights from the successes.

However, in this paper, we only analyze cities and towns within Massachusetts and do not consider the impact on other states. In our further study, we will take our focus on the influence of neighboring states to improve our method.
In addition, when calculating the transition probability matrix using GAT, due to the limited features of our data, we take the time series of confirmed cases every two weeks and obtain the transition probability of the comprehensive pandemic evolution for 41 weeks. Our further research will try to overcome the limited features of the data and train the network on the spatio-temporal dynamic graph to obtain the time-varying transition probability matrix.



\section*{Conflict of interest}
The authors declare that they have no conflict of interest.


\appendix[List of Massachusetts cities]

\begin{table*}[htbp]
\centering
\caption{List of Massachusetts cities}
\begin{tabular}{|llllllllll|}
\hline
\multicolumn{1}{|l|}{\textbf{ID}}  & \multicolumn{1}{l|}{\textbf{City/Town}} & \multicolumn{1}{l|}{\textbf{ID}}  & \multicolumn{1}{l|}{\textbf{City/Town}} & \multicolumn{1}{l|}{\textbf{ID}}  & \multicolumn{1}{l|}{\textbf{City/Town}} & \multicolumn{1}{l|}{\textbf{ID}}  & \multicolumn{1}{l|}{\textbf{City/Town}} & \multicolumn{1}{l|}{\textbf{ID}}  & \textbf{City/Town} \\ \hline
\multicolumn{1}{|l|}{\textbf{1}}   & \multicolumn{1}{l|}{Abington}           & \multicolumn{1}{l|}{\textbf{70}}  & \multicolumn{1}{l|}{Cummington}         & \multicolumn{1}{l|}{\textbf{142}} & \multicolumn{1}{l|}{Hull}               & \multicolumn{1}{l|}{\textbf{215}} & \multicolumn{1}{l|}{Northborough}       & \multicolumn{1}{l|}{\textbf{283}} & Stockbridge        \\ \hline
\multicolumn{1}{|l|}{\textbf{2}}   & \multicolumn{1}{l|}{Acton}              & \multicolumn{1}{l|}{\textbf{71}}  & \multicolumn{1}{l|}{Dalton}             & \multicolumn{1}{l|}{\textbf{143}} & \multicolumn{1}{l|}{Huntington}         & \multicolumn{1}{l|}{\textbf{216}} & \multicolumn{1}{l|}{Northbridge}        & \multicolumn{1}{l|}{\textbf{284}} & Stoneham           \\ \hline
\multicolumn{1}{|l|}{\textbf{3}}   & \multicolumn{1}{l|}{Acushnet}           & \multicolumn{1}{l|}{\textbf{72}}  & \multicolumn{1}{l|}{Danvers}            & \multicolumn{1}{l|}{\textbf{144}} & \multicolumn{1}{l|}{Ipswich}            & \multicolumn{1}{l|}{\textbf{217}} & \multicolumn{1}{l|}{Northfield}         & \multicolumn{1}{l|}{\textbf{285}} & Stoughton          \\ \hline
\multicolumn{1}{|l|}{\textbf{4}}   & \multicolumn{1}{l|}{Adams}              & \multicolumn{1}{l|}{\textbf{73}}  & \multicolumn{1}{l|}{Dartmouth}          & \multicolumn{1}{l|}{\textbf{145}} & \multicolumn{1}{l|}{Kingston}           & \multicolumn{1}{l|}{\textbf{218}} & \multicolumn{1}{l|}{Norton}             & \multicolumn{1}{l|}{\textbf{286}} & Stow               \\ \hline
\multicolumn{1}{|l|}{\textbf{5}}   & \multicolumn{1}{l|}{Agawam}             & \multicolumn{1}{l|}{\textbf{74}}  & \multicolumn{1}{l|}{Dedham}             & \multicolumn{1}{l|}{\textbf{146}} & \multicolumn{1}{l|}{Lakeville}          & \multicolumn{1}{l|}{\textbf{219}} & \multicolumn{1}{l|}{Norwell}            & \multicolumn{1}{l|}{\textbf{287}} & Sturbridge         \\ \hline
\multicolumn{1}{|l|}{\textbf{7}}   & \multicolumn{1}{l|}{Amesbury}           & \multicolumn{1}{l|}{\textbf{75}}  & \multicolumn{1}{l|}{Deerfield}          & \multicolumn{1}{l|}{\textbf{147}} & \multicolumn{1}{l|}{Lancaster}          & \multicolumn{1}{l|}{\textbf{220}} & \multicolumn{1}{l|}{Norwood}            & \multicolumn{1}{l|}{\textbf{288}} & Sudbury            \\ \hline
\multicolumn{1}{|l|}{\textbf{8}}   & \multicolumn{1}{l|}{Amherst}            & \multicolumn{1}{l|}{\textbf{76}}  & \multicolumn{1}{l|}{Dennis}             & \multicolumn{1}{l|}{\textbf{148}} & \multicolumn{1}{l|}{Lanesborough}       & \multicolumn{1}{l|}{\textbf{221}} & \multicolumn{1}{l|}{Oak Bluffs}         & \multicolumn{1}{l|}{\textbf{289}} & Sunderland         \\ \hline
\multicolumn{1}{|l|}{\textbf{9}}   & \multicolumn{1}{l|}{Andover}            & \multicolumn{1}{l|}{\textbf{77}}  & \multicolumn{1}{l|}{Dighton}            & \multicolumn{1}{l|}{\textbf{149}} & \multicolumn{1}{l|}{Lawrence}           & \multicolumn{1}{l|}{\textbf{222}} & \multicolumn{1}{l|}{Oakham}             & \multicolumn{1}{l|}{\textbf{290}} & Sutton             \\ \hline
\multicolumn{1}{|l|}{\textbf{10}}  & \multicolumn{1}{l|}{Aquinnah}           & \multicolumn{1}{l|}{\textbf{78}}  & \multicolumn{1}{l|}{Douglas}            & \multicolumn{1}{l|}{\textbf{150}} & \multicolumn{1}{l|}{Lee}                & \multicolumn{1}{l|}{\textbf{223}} & \multicolumn{1}{l|}{Orange}             & \multicolumn{1}{l|}{\textbf{291}} & Swampscott         \\ \hline
\multicolumn{1}{|l|}{\textbf{11}}  & \multicolumn{1}{l|}{Arlington}          & \multicolumn{1}{l|}{\textbf{80}}  & \multicolumn{1}{l|}{Dracut}             & \multicolumn{1}{l|}{\textbf{151}} & \multicolumn{1}{l|}{Leicester}          & \multicolumn{1}{l|}{\textbf{224}} & \multicolumn{1}{l|}{Orleans}            & \multicolumn{1}{l|}{\textbf{292}} & Swansea            \\ \hline
\multicolumn{1}{|l|}{\textbf{12}}  & \multicolumn{1}{l|}{Ashburnham}         & \multicolumn{1}{l|}{\textbf{81}}  & \multicolumn{1}{l|}{Dudley}             & \multicolumn{1}{l|}{\textbf{152}} & \multicolumn{1}{l|}{Lenox}              & \multicolumn{1}{l|}{\textbf{225}} & \multicolumn{1}{l|}{Otis}               & \multicolumn{1}{l|}{\textbf{293}} & Taunton            \\ \hline
\multicolumn{1}{|l|}{\textbf{13}}  & \multicolumn{1}{l|}{Ashby}              & \multicolumn{1}{l|}{\textbf{82}}  & \multicolumn{1}{l|}{Dunstable}          & \multicolumn{1}{l|}{\textbf{153}} & \multicolumn{1}{l|}{Leominster}         & \multicolumn{1}{l|}{\textbf{226}} & \multicolumn{1}{l|}{Oxford}             & \multicolumn{1}{l|}{\textbf{294}} & Templeton          \\ \hline
\multicolumn{1}{|l|}{\textbf{14}}  & \multicolumn{1}{l|}{Ashfield}           & \multicolumn{1}{l|}{\textbf{83}}  & \multicolumn{1}{l|}{Duxbury}            & \multicolumn{1}{l|}{\textbf{154}} & \multicolumn{1}{l|}{Leverett}           & \multicolumn{1}{l|}{\textbf{227}} & \multicolumn{1}{l|}{Palmer}             & \multicolumn{1}{l|}{\textbf{295}} & Tewksbury          \\ \hline
\multicolumn{1}{|l|}{\textbf{15}}  & \multicolumn{1}{l|}{Ashland}            & \multicolumn{1}{l|}{\textbf{84}}  & \multicolumn{1}{l|}{East Bridgewater}   & \multicolumn{1}{l|}{\textbf{155}} & \multicolumn{1}{l|}{Lexington}          & \multicolumn{1}{l|}{\textbf{228}} & \multicolumn{1}{l|}{Paxton}             & \multicolumn{1}{l|}{\textbf{296}} & Tisbury            \\ \hline
\multicolumn{1}{|l|}{\textbf{16}}  & \multicolumn{1}{l|}{Athol}              & \multicolumn{1}{l|}{\textbf{85}}  & \multicolumn{1}{l|}{East Brookfield}    & \multicolumn{1}{l|}{\textbf{157}} & \multicolumn{1}{l|}{Lincoln}            & \multicolumn{1}{l|}{\textbf{229}} & \multicolumn{1}{l|}{Peabody}            & \multicolumn{1}{l|}{\textbf{297}} & Tolland            \\ \hline
\multicolumn{1}{|l|}{\textbf{17}}  & \multicolumn{1}{l|}{Attleboro}          & \multicolumn{1}{l|}{\textbf{86}}  & \multicolumn{1}{l|}{East Longmeadow}    & \multicolumn{1}{l|}{\textbf{158}} & \multicolumn{1}{l|}{Littleton}          & \multicolumn{1}{l|}{\textbf{230}} & \multicolumn{1}{l|}{Pelham}             & \multicolumn{1}{l|}{\textbf{298}} & Topsfield          \\ \hline
\multicolumn{1}{|l|}{\textbf{18}}  & \multicolumn{1}{l|}{Auburn}             & \multicolumn{1}{l|}{\textbf{87}}  & \multicolumn{1}{l|}{Eastham}            & \multicolumn{1}{l|}{\textbf{159}} & \multicolumn{1}{l|}{Longmeadow}         & \multicolumn{1}{l|}{\textbf{231}} & \multicolumn{1}{l|}{Pembroke}           & \multicolumn{1}{l|}{\textbf{299}} & Townsend           \\ \hline
\multicolumn{1}{|l|}{\textbf{19}}  & \multicolumn{1}{l|}{Avon}               & \multicolumn{1}{l|}{\textbf{88}}  & \multicolumn{1}{l|}{Easthampton}        & \multicolumn{1}{l|}{\textbf{160}} & \multicolumn{1}{l|}{Lowell}             & \multicolumn{1}{l|}{\textbf{232}} & \multicolumn{1}{l|}{Pepperell}          & \multicolumn{1}{l|}{\textbf{300}} & Truro              \\ \hline
\multicolumn{1}{|l|}{\textbf{20}}  & \multicolumn{1}{l|}{Ayer}               & \multicolumn{1}{l|}{\textbf{89}}  & \multicolumn{1}{l|}{Easton}             & \multicolumn{1}{l|}{\textbf{161}} & \multicolumn{1}{l|}{Ludlow}             & \multicolumn{1}{l|}{\textbf{233}} & \multicolumn{1}{l|}{Peru}               & \multicolumn{1}{l|}{\textbf{301}} & Tyngsborough       \\ \hline
\multicolumn{1}{|l|}{\textbf{21}}  & \multicolumn{1}{l|}{Barnstable}         & \multicolumn{1}{l|}{\textbf{90}}  & \multicolumn{1}{l|}{Edgartown}          & \multicolumn{1}{l|}{\textbf{162}} & \multicolumn{1}{l|}{Lunenburg}          & \multicolumn{1}{l|}{\textbf{234}} & \multicolumn{1}{l|}{Petersham}          & \multicolumn{1}{l|}{\textbf{303}} & Upton              \\ \hline
\multicolumn{1}{|l|}{\textbf{22}}  & \multicolumn{1}{l|}{Barre}              & \multicolumn{1}{l|}{\textbf{91}}  & \multicolumn{1}{l|}{Egremont}           & \multicolumn{1}{l|}{\textbf{163}} & \multicolumn{1}{l|}{Lynn}               & \multicolumn{1}{l|}{\textbf{235}} & \multicolumn{1}{l|}{Phillipston}        & \multicolumn{1}{l|}{\textbf{304}} & Uxbridge           \\ \hline
\multicolumn{1}{|l|}{\textbf{23}}  & \multicolumn{1}{l|}{Becket}             & \multicolumn{1}{l|}{\textbf{92}}  & \multicolumn{1}{l|}{Erving}             & \multicolumn{1}{l|}{\textbf{164}} & \multicolumn{1}{l|}{Lynnfield}          & \multicolumn{1}{l|}{\textbf{236}} & \multicolumn{1}{l|}{Pittsfield}         & \multicolumn{1}{l|}{\textbf{305}} & Wakefield          \\ \hline
\multicolumn{1}{|l|}{\textbf{24}}  & \multicolumn{1}{l|}{Bedford}            & \multicolumn{1}{l|}{\textbf{93}}  & \multicolumn{1}{l|}{Essex}              & \multicolumn{1}{l|}{\textbf{165}} & \multicolumn{1}{l|}{Malden}             & \multicolumn{1}{l|}{\textbf{237}} & \multicolumn{1}{l|}{Plainfield}         & \multicolumn{1}{l|}{\textbf{306}} & Wales              \\ \hline
\multicolumn{1}{|l|}{\textbf{25}}  & \multicolumn{1}{l|}{Belchertown}        & \multicolumn{1}{l|}{\textbf{94}}  & \multicolumn{1}{l|}{Everett}            & \multicolumn{1}{l|}{\textbf{166}} & \multicolumn{1}{l|}{Manchester}         & \multicolumn{1}{l|}{\textbf{238}} & \multicolumn{1}{l|}{Plainville}         & \multicolumn{1}{l|}{\textbf{307}} & Walpole            \\ \hline
\multicolumn{1}{|l|}{\textbf{26}}  & \multicolumn{1}{l|}{Bellingham}         & \multicolumn{1}{l|}{\textbf{95}}  & \multicolumn{1}{l|}{Fairhaven}          & \multicolumn{1}{l|}{\textbf{167}} & \multicolumn{1}{l|}{Mansfield}          & \multicolumn{1}{l|}{\textbf{239}} & \multicolumn{1}{l|}{Plymouth}           & \multicolumn{1}{l|}{\textbf{308}} & Waltham            \\ \hline
\multicolumn{1}{|l|}{\textbf{27}}  & \multicolumn{1}{l|}{Belmont}            & \multicolumn{1}{l|}{\textbf{96}}  & \multicolumn{1}{l|}{Fall River}         & \multicolumn{1}{l|}{\textbf{168}} & \multicolumn{1}{l|}{Marblehead}         & \multicolumn{1}{l|}{\textbf{240}} & \multicolumn{1}{l|}{Plympton}           & \multicolumn{1}{l|}{\textbf{309}} & Ware               \\ \hline
\multicolumn{1}{|l|}{\textbf{28}}  & \multicolumn{1}{l|}{Berkley}            & \multicolumn{1}{l|}{\textbf{97}}  & \multicolumn{1}{l|}{Falmouth}           & \multicolumn{1}{l|}{\textbf{169}} & \multicolumn{1}{l|}{Marion}             & \multicolumn{1}{l|}{\textbf{241}} & \multicolumn{1}{l|}{Princeton}          & \multicolumn{1}{l|}{\textbf{310}} & Wareham            \\ \hline
\multicolumn{1}{|l|}{\textbf{29}}  & \multicolumn{1}{l|}{Berlin}             & \multicolumn{1}{l|}{\textbf{98}}  & \multicolumn{1}{l|}{Fitchburg}          & \multicolumn{1}{l|}{\textbf{170}} & \multicolumn{1}{l|}{Marlborough}        & \multicolumn{1}{l|}{\textbf{242}} & \multicolumn{1}{l|}{Provincetown}       & \multicolumn{1}{l|}{\textbf{311}} & Warren             \\ \hline
\multicolumn{1}{|l|}{\textbf{30}}  & \multicolumn{1}{l|}{Bernardston}        & \multicolumn{1}{l|}{\textbf{99}}  & \multicolumn{1}{l|}{Florida}            & \multicolumn{1}{l|}{\textbf{171}} & \multicolumn{1}{l|}{Marshfield}         & \multicolumn{1}{l|}{\textbf{243}} & \multicolumn{1}{l|}{Quincy}             & \multicolumn{1}{l|}{\textbf{312}} & Warwick            \\ \hline
\multicolumn{1}{|l|}{\textbf{31}}  & \multicolumn{1}{l|}{Beverly}            & \multicolumn{1}{l|}{\textbf{100}} & \multicolumn{1}{l|}{Foxborough}         & \multicolumn{1}{l|}{\textbf{172}} & \multicolumn{1}{l|}{Mashpee}            & \multicolumn{1}{l|}{\textbf{244}} & \multicolumn{1}{l|}{Randolph}           & \multicolumn{1}{l|}{\textbf{313}} & Washington         \\ \hline
\multicolumn{1}{|l|}{\textbf{32}}  & \multicolumn{1}{l|}{Billerica}          & \multicolumn{1}{l|}{\textbf{101}} & \multicolumn{1}{l|}{Framingham}         & \multicolumn{1}{l|}{\textbf{173}} & \multicolumn{1}{l|}{Mattapoisett}       & \multicolumn{1}{l|}{\textbf{245}} & \multicolumn{1}{l|}{Raynham}            & \multicolumn{1}{l|}{\textbf{314}} & Watertown          \\ \hline
\multicolumn{1}{|l|}{\textbf{33}}  & \multicolumn{1}{l|}{Blackstone}         & \multicolumn{1}{l|}{\textbf{102}} & \multicolumn{1}{l|}{Franklin}           & \multicolumn{1}{l|}{\textbf{174}} & \multicolumn{1}{l|}{Maynard}            & \multicolumn{1}{l|}{\textbf{246}} & \multicolumn{1}{l|}{Reading}            & \multicolumn{1}{l|}{\textbf{315}} & Wayland            \\ \hline
\multicolumn{1}{|l|}{\textbf{34}}  & \multicolumn{1}{l|}{Blandford}          & \multicolumn{1}{l|}{\textbf{103}} & \multicolumn{1}{l|}{Freetown}           & \multicolumn{1}{l|}{\textbf{175}} & \multicolumn{1}{l|}{Medfield}           & \multicolumn{1}{l|}{\textbf{247}} & \multicolumn{1}{l|}{Rehoboth}           & \multicolumn{1}{l|}{\textbf{316}} & Webster            \\ \hline
\multicolumn{1}{|l|}{\textbf{35}}  & \multicolumn{1}{l|}{Bolton}             & \multicolumn{1}{l|}{\textbf{104}} & \multicolumn{1}{l|}{Gardner}            & \multicolumn{1}{l|}{\textbf{176}} & \multicolumn{1}{l|}{Medford}            & \multicolumn{1}{l|}{\textbf{248}} & \multicolumn{1}{l|}{Revere}             & \multicolumn{1}{l|}{\textbf{317}} & Wellesley          \\ \hline
\multicolumn{1}{|l|}{\textbf{36}}  & \multicolumn{1}{l|}{Boston}             & \multicolumn{1}{l|}{\textbf{105}} & \multicolumn{1}{l|}{Georgetown}         & \multicolumn{1}{l|}{\textbf{177}} & \multicolumn{1}{l|}{Medway}             & \multicolumn{1}{l|}{\textbf{249}} & \multicolumn{1}{l|}{Richmond}           & \multicolumn{1}{l|}{\textbf{318}} & Wellfleet          \\ \hline
\multicolumn{1}{|l|}{\textbf{37}}  & \multicolumn{1}{l|}{Bourne}             & \multicolumn{1}{l|}{\textbf{106}} & \multicolumn{1}{l|}{Gill}               & \multicolumn{1}{l|}{\textbf{178}} & \multicolumn{1}{l|}{Melrose}            & \multicolumn{1}{l|}{\textbf{250}} & \multicolumn{1}{l|}{Rochester}          & \multicolumn{1}{l|}{\textbf{320}} & Wenham             \\ \hline
\multicolumn{1}{|l|}{\textbf{38}}  & \multicolumn{1}{l|}{Boxborough}         & \multicolumn{1}{l|}{\textbf{107}} & \multicolumn{1}{l|}{Gloucester}         & \multicolumn{1}{l|}{\textbf{179}} & \multicolumn{1}{l|}{Mendon}             & \multicolumn{1}{l|}{\textbf{251}} & \multicolumn{1}{l|}{Rockland}           & \multicolumn{1}{l|}{\textbf{321}} & West Boylston      \\ \hline
\multicolumn{1}{|l|}{\textbf{39}}  & \multicolumn{1}{l|}{Boxford}            & \multicolumn{1}{l|}{\textbf{108}} & \multicolumn{1}{l|}{Goshen}             & \multicolumn{1}{l|}{\textbf{180}} & \multicolumn{1}{l|}{Merrimac}           & \multicolumn{1}{l|}{\textbf{252}} & \multicolumn{1}{l|}{Rockport}           & \multicolumn{1}{l|}{\textbf{322}} & West Bridgewater   \\ \hline
\multicolumn{1}{|l|}{\textbf{40}}  & \multicolumn{1}{l|}{Boylston}           & \multicolumn{1}{l|}{\textbf{110}} & \multicolumn{1}{l|}{Grafton}            & \multicolumn{1}{l|}{\textbf{181}} & \multicolumn{1}{l|}{Methuen}            & \multicolumn{1}{l|}{\textbf{254}} & \multicolumn{1}{l|}{Rowley}             & \multicolumn{1}{l|}{\textbf{323}} & West Brookfield    \\ \hline
\multicolumn{1}{|l|}{\textbf{41}}  & \multicolumn{1}{l|}{Braintree}          & \multicolumn{1}{l|}{\textbf{111}} & \multicolumn{1}{l|}{Granby}             & \multicolumn{1}{l|}{\textbf{182}} & \multicolumn{1}{l|}{Middleborough}      & \multicolumn{1}{l|}{\textbf{255}} & \multicolumn{1}{l|}{Royalston}          & \multicolumn{1}{l|}{\textbf{324}} & West Newbury       \\ \hline
\multicolumn{1}{|l|}{\textbf{42}}  & \multicolumn{1}{l|}{Brewster}           & \multicolumn{1}{l|}{\textbf{112}} & \multicolumn{1}{l|}{Granville}          & \multicolumn{1}{l|}{\textbf{184}} & \multicolumn{1}{l|}{Middleton}          & \multicolumn{1}{l|}{\textbf{256}} & \multicolumn{1}{l|}{Russell}            & \multicolumn{1}{l|}{\textbf{325}} & West Springfield   \\ \hline
\multicolumn{1}{|l|}{\textbf{43}}  & \multicolumn{1}{l|}{Bridgewater}        & \multicolumn{1}{l|}{\textbf{113}} & \multicolumn{1}{l|}{Great Barrington}   & \multicolumn{1}{l|}{\textbf{185}} & \multicolumn{1}{l|}{Milford}            & \multicolumn{1}{l|}{\textbf{257}} & \multicolumn{1}{l|}{Rutland}            & \multicolumn{1}{l|}{\textbf{326}} & West Stockbridge   \\ \hline
\multicolumn{1}{|l|}{\textbf{44}}  & \multicolumn{1}{l|}{Brimfield}          & \multicolumn{1}{l|}{\textbf{114}} & \multicolumn{1}{l|}{Greenfield}         & \multicolumn{1}{l|}{\textbf{186}} & \multicolumn{1}{l|}{Millbury}           & \multicolumn{1}{l|}{\textbf{258}} & \multicolumn{1}{l|}{Salem}              & \multicolumn{1}{l|}{\textbf{327}} & West Tisbury       \\ \hline
\multicolumn{1}{|l|}{\textbf{45}}  & \multicolumn{1}{l|}{Brockton}           & \multicolumn{1}{l|}{\textbf{115}} & \multicolumn{1}{l|}{Groton}             & \multicolumn{1}{l|}{\textbf{187}} & \multicolumn{1}{l|}{Millis}             & \multicolumn{1}{l|}{\textbf{259}} & \multicolumn{1}{l|}{Salisbury}          & \multicolumn{1}{l|}{\textbf{328}} & Westborough        \\ \hline
\multicolumn{1}{|l|}{\textbf{46}}  & \multicolumn{1}{l|}{Brookfield}         & \multicolumn{1}{l|}{\textbf{116}} & \multicolumn{1}{l|}{Groveland}          & \multicolumn{1}{l|}{\textbf{188}} & \multicolumn{1}{l|}{Millville}          & \multicolumn{1}{l|}{\textbf{260}} & \multicolumn{1}{l|}{Sandisfield}        & \multicolumn{1}{l|}{\textbf{329}} & Westfield          \\ \hline
\multicolumn{1}{|l|}{\textbf{47}}  & \multicolumn{1}{l|}{Brookline}          & \multicolumn{1}{l|}{\textbf{117}} & \multicolumn{1}{l|}{Hadley}             & \multicolumn{1}{l|}{\textbf{189}} & \multicolumn{1}{l|}{Milton}             & \multicolumn{1}{l|}{\textbf{261}} & \multicolumn{1}{l|}{Sandwich}           & \multicolumn{1}{l|}{\textbf{330}} & Westford           \\ \hline
\multicolumn{1}{|l|}{\textbf{48}}  & \multicolumn{1}{l|}{Buckland}           & \multicolumn{1}{l|}{\textbf{118}} & \multicolumn{1}{l|}{Halifax}            & \multicolumn{1}{l|}{\textbf{191}} & \multicolumn{1}{l|}{Monson}             & \multicolumn{1}{l|}{\textbf{262}} & \multicolumn{1}{l|}{Saugus}             & \multicolumn{1}{l|}{\textbf{331}} & Westhampton        \\ \hline
\multicolumn{1}{|l|}{\textbf{49}}  & \multicolumn{1}{l|}{Burlington}         & \multicolumn{1}{l|}{\textbf{119}} & \multicolumn{1}{l|}{Hamilton}           & \multicolumn{1}{l|}{\textbf{192}} & \multicolumn{1}{l|}{Montague}           & \multicolumn{1}{l|}{\textbf{263}} & \multicolumn{1}{l|}{Savoy}              & \multicolumn{1}{l|}{\textbf{332}} & Westminster        \\ \hline
\multicolumn{1}{|l|}{\textbf{50}}  & \multicolumn{1}{l|}{Cambridge}          & \multicolumn{1}{l|}{\textbf{121}} & \multicolumn{1}{l|}{Hancock}            & \multicolumn{1}{l|}{\textbf{193}} & \multicolumn{1}{l|}{Monterey}           & \multicolumn{1}{l|}{\textbf{264}} & \multicolumn{1}{l|}{Scituate}           & \multicolumn{1}{l|}{\textbf{333}} & Weston             \\ \hline
\multicolumn{1}{|l|}{\textbf{51}}  & \multicolumn{1}{l|}{Canton}             & \multicolumn{1}{l|}{\textbf{122}} & \multicolumn{1}{l|}{Hanover}            & \multicolumn{1}{l|}{\textbf{196}} & \multicolumn{1}{l|}{Nahant}             & \multicolumn{1}{l|}{\textbf{265}} & \multicolumn{1}{l|}{Seekonk}            & \multicolumn{1}{l|}{\textbf{334}} & Westport           \\ \hline
\multicolumn{1}{|l|}{\textbf{52}}  & \multicolumn{1}{l|}{Carlisle}           & \multicolumn{1}{l|}{\textbf{123}} & \multicolumn{1}{l|}{Hanson}             & \multicolumn{1}{l|}{\textbf{197}} & \multicolumn{1}{l|}{Nantucket}          & \multicolumn{1}{l|}{\textbf{266}} & \multicolumn{1}{l|}{Sharon}             & \multicolumn{1}{l|}{\textbf{335}} & Westwood           \\ \hline
\multicolumn{1}{|l|}{\textbf{53}}  & \multicolumn{1}{l|}{Carver}             & \multicolumn{1}{l|}{\textbf{124}} & \multicolumn{1}{l|}{Hardwick}           & \multicolumn{1}{l|}{\textbf{198}} & \multicolumn{1}{l|}{Natick}             & \multicolumn{1}{l|}{\textbf{267}} & \multicolumn{1}{l|}{Sheffield}          & \multicolumn{1}{l|}{\textbf{336}} & Weymouth           \\ \hline
\multicolumn{1}{|l|}{\textbf{54}}  & \multicolumn{1}{l|}{Charlemont}         & \multicolumn{1}{l|}{\textbf{125}} & \multicolumn{1}{l|}{Harvard}            & \multicolumn{1}{l|}{\textbf{199}} & \multicolumn{1}{l|}{Needham}            & \multicolumn{1}{l|}{\textbf{268}} & \multicolumn{1}{l|}{Shelburne}          & \multicolumn{1}{l|}{\textbf{337}} & Whately            \\ \hline
\multicolumn{1}{|l|}{\textbf{55}}  & \multicolumn{1}{l|}{Charlton}           & \multicolumn{1}{l|}{\textbf{126}} & \multicolumn{1}{l|}{Harwich}            & \multicolumn{1}{l|}{\textbf{200}} & \multicolumn{1}{l|}{New Ashford}        & \multicolumn{1}{l|}{\textbf{269}} & \multicolumn{1}{l|}{Sherborn}           & \multicolumn{1}{l|}{\textbf{338}} & Whitman            \\ \hline
\multicolumn{1}{|l|}{\textbf{56}}  & \multicolumn{1}{l|}{Chatham}            & \multicolumn{1}{l|}{\textbf{127}} & \multicolumn{1}{l|}{Hatfield}           & \multicolumn{1}{l|}{\textbf{201}} & \multicolumn{1}{l|}{New Bedford}        & \multicolumn{1}{l|}{\textbf{270}} & \multicolumn{1}{l|}{Shirley}            & \multicolumn{1}{l|}{\textbf{339}} & Wilbraham          \\ \hline
\multicolumn{1}{|l|}{\textbf{57}}  & \multicolumn{1}{l|}{Chelmsford}         & \multicolumn{1}{l|}{\textbf{128}} & \multicolumn{1}{l|}{Haverhill}          & \multicolumn{1}{l|}{\textbf{202}} & \multicolumn{1}{l|}{New Braintree}      & \multicolumn{1}{l|}{\textbf{271}} & \multicolumn{1}{l|}{Shrewsbury}         & \multicolumn{1}{l|}{\textbf{340}} & Williamsburg       \\ \hline
\multicolumn{1}{|l|}{\textbf{58}}  & \multicolumn{1}{l|}{Chelsea}            & \multicolumn{1}{l|}{\textbf{129}} & \multicolumn{1}{l|}{Hawley}             & \multicolumn{1}{l|}{\textbf{203}} & \multicolumn{1}{l|}{New Marlborough}    & \multicolumn{1}{l|}{\textbf{272}} & \multicolumn{1}{l|}{Shutesbury}         & \multicolumn{1}{l|}{\textbf{341}} & Williamstown       \\ \hline
\multicolumn{1}{|l|}{\textbf{59}}  & \multicolumn{1}{l|}{Cheshire}           & \multicolumn{1}{l|}{\textbf{130}} & \multicolumn{1}{l|}{Heath}              & \multicolumn{1}{l|}{\textbf{204}} & \multicolumn{1}{l|}{New Salem}          & \multicolumn{1}{l|}{\textbf{273}} & \multicolumn{1}{l|}{Somerset}           & \multicolumn{1}{l|}{\textbf{342}} & Wilmington         \\ \hline
\multicolumn{1}{|l|}{\textbf{60}}  & \multicolumn{1}{l|}{Chester}            & \multicolumn{1}{l|}{\textbf{131}} & \multicolumn{1}{l|}{Hingham}            & \multicolumn{1}{l|}{\textbf{205}} & \multicolumn{1}{l|}{Newbury}            & \multicolumn{1}{l|}{\textbf{274}} & \multicolumn{1}{l|}{Somerville}         & \multicolumn{1}{l|}{\textbf{343}} & Winchendon         \\ \hline
\multicolumn{1}{|l|}{\textbf{61}}  & \multicolumn{1}{l|}{Chesterfield}       & \multicolumn{1}{l|}{\textbf{132}} & \multicolumn{1}{l|}{Hinsdale}           & \multicolumn{1}{l|}{\textbf{206}} & \multicolumn{1}{l|}{Newburyport}        & \multicolumn{1}{l|}{\textbf{275}} & \multicolumn{1}{l|}{South Hadley}       & \multicolumn{1}{l|}{\textbf{344}} & Winchester         \\ \hline
\multicolumn{1}{|l|}{\textbf{62}}  & \multicolumn{1}{l|}{Chicopee}           & \multicolumn{1}{l|}{\textbf{133}} & \multicolumn{1}{l|}{Holbrook}           & \multicolumn{1}{l|}{\textbf{207}} & \multicolumn{1}{l|}{Newton}             & \multicolumn{1}{l|}{\textbf{276}} & \multicolumn{1}{l|}{Southampton}        & \multicolumn{1}{l|}{\textbf{345}} & Windsor            \\ \hline
\multicolumn{1}{|l|}{\textbf{63}}  & \multicolumn{1}{l|}{Chilmark}           & \multicolumn{1}{l|}{\textbf{134}} & \multicolumn{1}{l|}{Holden}             & \multicolumn{1}{l|}{\textbf{208}} & \multicolumn{1}{l|}{Norfolk}            & \multicolumn{1}{l|}{\textbf{277}} & \multicolumn{1}{l|}{Southborough}       & \multicolumn{1}{l|}{\textbf{346}} & Winthrop           \\ \hline
\multicolumn{1}{|l|}{\textbf{64}}  & \multicolumn{1}{l|}{Clarksburg}         & \multicolumn{1}{l|}{\textbf{136}} & \multicolumn{1}{l|}{Holliston}          & \multicolumn{1}{l|}{\textbf{209}} & \multicolumn{1}{l|}{North Adams}        & \multicolumn{1}{l|}{\textbf{278}} & \multicolumn{1}{l|}{Southbridge}        & \multicolumn{1}{l|}{\textbf{347}} & Woburn             \\ \hline
\multicolumn{1}{|l|}{\textbf{65}}  & \multicolumn{1}{l|}{Clinton}            & \multicolumn{1}{l|}{\textbf{137}} & \multicolumn{1}{l|}{Holyoke}            & \multicolumn{1}{l|}{\textbf{210}} & \multicolumn{1}{l|}{North Andover}      & \multicolumn{1}{l|}{\textbf{279}} & \multicolumn{1}{l|}{Southwick}          & \multicolumn{1}{l|}{\textbf{348}} & Worcester          \\ \hline
\multicolumn{1}{|l|}{\textbf{66}}  & \multicolumn{1}{l|}{Cohasset}           & \multicolumn{1}{l|}{\textbf{138}} & \multicolumn{1}{l|}{Hopedale}           & \multicolumn{1}{l|}{\textbf{211}} & \multicolumn{1}{l|}{North Attleborough} & \multicolumn{1}{l|}{\textbf{280}} & \multicolumn{1}{l|}{Spencer}            & \multicolumn{1}{l|}{\textbf{349}} & Worthington        \\ \hline
\multicolumn{1}{|l|}{\textbf{67}}  & \multicolumn{1}{l|}{Colrain}            & \multicolumn{1}{l|}{\textbf{139}} & \multicolumn{1}{l|}{Hopkinton}          & \multicolumn{1}{l|}{\textbf{212}} & \multicolumn{1}{l|}{North Brookfield}   & \multicolumn{1}{l|}{\textbf{281}} & \multicolumn{1}{l|}{Springfield}        & \multicolumn{1}{l|}{\textbf{350}} & Wrentham           \\ \hline
\multicolumn{1}{|l|}{\textbf{68}}  & \multicolumn{1}{l|}{Concord}            & \multicolumn{1}{l|}{\textbf{140}} & \multicolumn{1}{l|}{Hubbardston}        & \multicolumn{1}{l|}{\textbf{213}} & \multicolumn{1}{l|}{North Reading}      & \multicolumn{1}{l|}{\textbf{282}} & \multicolumn{1}{l|}{Sterling}           & \multicolumn{1}{l|}{\textbf{351}} & Yarmouth           \\ \hline
\multicolumn{1}{|l|}{\textbf{69}}  & \multicolumn{1}{l|}{Conway}             & \multicolumn{1}{l|}{\textbf{141}} & \multicolumn{1}{l|}{Hudson}             & \multicolumn{1}{l|}{\textbf{214}} & \multicolumn{1}{l|}{Northampton}        & \multicolumn{1}{l|}{\textbf{}}    & \multicolumn{1}{l|}{\textbf{}}          & \multicolumn{1}{l|}{\textbf{}}    & \textbf{}          \\ \hline
\multicolumn{10}{|l|}{\textbf{Thirteen isolated cities}}                                                                                                                                                                                                                                                                                                                        \\ \hline
\multicolumn{1}{|l|}{\textbf{6}}   & \multicolumn{1}{l|}{Alford}             & \multicolumn{1}{l|}{\textbf{79}}  & \multicolumn{1}{l|}{Dover}              & \multicolumn{1}{l|}{\textbf{109}} & \multicolumn{1}{l|}{Gosnold}            & \multicolumn{1}{l|}{\textbf{120}} & \multicolumn{1}{l|}{Hampden}            & \multicolumn{1}{l|}{\textbf{135}} & Holland            \\ \hline
\multicolumn{1}{|l|}{\textbf{156}}          & \multicolumn{1}{l|}{Leyden}             & \multicolumn{1}{l|}{\textbf{183}} & \multicolumn{1}{l|}{Middlefield}        & \multicolumn{1}{l|}{\textbf{190}} & \multicolumn{1}{l|}{Monroe}             & \multicolumn{1}{l|}{\textbf{194}} & \multicolumn{1}{l|}{Montgomery}         & \multicolumn{1}{l|}{\textbf{195}} & Mount Washington   \\ \hline
\multicolumn{1}{|l|}{\textbf{253}} & \multicolumn{1}{l|}{Rowe}               & \multicolumn{1}{l|}{\textbf{302}} & \multicolumn{1}{l|}{Tyringham}          & \multicolumn{1}{l|}{\textbf{319}} & \multicolumn{1}{l|}{Wendell}            & \multicolumn{1}{l|}{\textbf{}}    & \multicolumn{1}{l|}{\textbf{}}          & \multicolumn{1}{l|}{\textbf{}}    & \textbf{}          \\ \hline
\end{tabular}
\end{table*}


\bibliographystyle{IEEEtran}

\bibliography{refs} 

\begin{thebibliography}{10}
\providecommand{\url}[1]{#1}
\csname url@samestyle\endcsname
\providecommand{\newblock}{\relax}
\providecommand{\bibinfo}[2]{#2}
\providecommand{\BIBentrySTDinterwordspacing}{\spaceskip=0pt\relax}
\providecommand{\BIBentryALTinterwordstretchfactor}{4}
\providecommand{\BIBentryALTinterwordspacing}{\spaceskip=\fontdimen2\font plus
\BIBentryALTinterwordstretchfactor\fontdimen3\font minus
  \fontdimen4\font\relax}
\providecommand{\BIBforeignlanguage}[2]{{%
\expandafter\ifx\csname l@#1\endcsname\relax
\typeout{** WARNING: IEEEtran.bst: No hyphenation pattern has been}%
\typeout{** loaded for the language `#1'. Using the pattern for}%
\typeout{** the default language instead.}%
\else
\language=\csname l@#1\endcsname
\fi
#2}}
\providecommand{\BIBdecl}{\relax}
\BIBdecl

\bibitem{WHO2020}
{World Health Organization}, ``{WHO} director-general's opening remarks at the
  media briefing on {COVID-19} - 11 march 2020,''
  \url{https://www.who.int/director-general/speeches/detail/who-director-general-s-opening-remarks-at-the-media-briefing-on-covid-19---11-march-2020},
  Mar. 11, 2020 [Online].

\bibitem{JHU2021}
{Johns Hopkins Coronavirus Resource Center}, ``{COVID-19} dashboard,''
  \url{https: //coronavirus.jhu.edu/map.html}, Dec. 1, 2021 [Online].

\bibitem{gao2021transmission}
Q.~Gao, J.~Zhuang, T.~Wu, and H.~Shen, ``Transmission dynamics and quarantine
  control of covid-19 in cluster community: A new transmission-quarantine model
  with case study for diamond princess,'' \emph{Mathematical Models and Methods
  in Applied Sciences}, vol.~31, no.~03, pp. 619--648, 2021.

\bibitem{church2021}
K.~E. Church, ``Analysis of pandemic closing-reopening cycles using rigorous
  homotopy continuation: a case study with montreal {COVID-19} data,''
  \emph{SIAM Journal on Applied Dynamical Systems}, vol.~20, no.~2, pp.
  745--783, 2021.

\bibitem{neves2020}
A.~G. Neves and G.~Guerrero, ``Predicting the evolution of the {COVID-19}
  epidemic with the a-sir model: Lombardy, italy and sao paulo state, brazil,''
  \emph{Physica D: Nonlinear Phenomena}, vol. 413, p. 132693, 2020.

\bibitem{ng2020}
K.~Y. Ng and M.~M. Gui, ``{COVID-19}: development of a robust mathematical
  model and simulation package with consideration for ageing population and
  time delay for control action and resusceptibility,'' \emph{Physica D:
  Nonlinear Phenomena}, vol. 411, p. 132599, 2020.

\bibitem{miranda2021}
J.~G.~V. Miranda, M.~S. Silva, J.~G. Bertolino, R.~N. Vasconcelos, E.~C.~B.
  Cambui, M.~L.~V. Ara{\'u}jo, H.~Saba, D.~P. Costa, S.~G. Duverger, M.~T.
  de~Oliveira \emph{et~al.}, ``Scaling effect in {COVID-19} spreading: The role
  of heterogeneity in a hybrid ode-network model with restrictions on the
  inter-cities flow,'' \emph{Physica D: Nonlinear Phenomena}, vol. 415, p.
  132792, 2021.

\bibitem{tang2021}
F.~Tang, Y.~Feng, H.~Chiheb, and J.~Fan, ``The interplay of demographic
  variables and social distancing scores in deep prediction of {US} {COVID-19}
  cases,'' \emph{Journal of the American Statistical Association}, vol. 116,
  no. 534, pp. 492--506, 2021.

\bibitem{Melin2020}
P.~Melin, J.~C. Monica, D.~Sanchez, and O.~Castillo, ``Analysis of spatial
  spread relationships of coronavirus ({COVID-19}) pandemic in the world using
  self organizing maps,'' \emph{Chaos Solitons and Fractals}, vol. 138, p.
  109917, 2020.

\bibitem{TatDat2020}
T.~Tat~Dat, P.~Fr{\'e}d{\'e}ric, N.~T. Hang, M.~Jules, N.~Duc~Thang,
  C.~Piffault, R.~Willy, F.~Susely, H.~V. L{\^e}, W.~Tuschmann \emph{et~al.},
  ``Epidemic dynamics via wavelet theory and machine learning with applications
  to {COVID-19},'' \emph{Biology}, vol.~9, no.~12, p. 477, 2020.

\bibitem{LiMateos2021}
Y.~Li and G.~Mateos, ``Graph frequency analysis of {COVID-19} incidence to
  identify county-level contagion patterns in the {United States},'' in
  \emph{ICASSP 2021-2021 IEEE International Conference on Acoustics, Speech and
  Signal Processing (ICASSP)}.\hskip 1em plus 0.5em minus 0.4em\relax IEEE,
  2021, pp. 3230--3234.

\bibitem{gao2021}
J.~Gao, R.~Sharma, C.~Qian, L.~M. Glass, J.~Spaeder, J.~Romberg, J.~Sun, and
  C.~Xiao, ``{STAN}: spatio-temporal attention network for pandemic prediction
  using real-world evidence,'' \emph{Journal of the American Medical
  Informatics Association}, vol.~28, no.~4, pp. 733--743, 2021.

\bibitem{kapoor2020}
A.~Kapoor, X.~Ben, L.~Liu, B.~Perozzi, M.~Barnes, M.~Blais, and S.~O'Banion,
  ``Examining {COVID-19} forecasting using spatio-temporal graph neural
  networks,'' \emph{arXiv preprint arXiv:2007.03113}, 2020.

\bibitem{Velikovi2018}
P.~Veli{\v{c}}kovi{\'c}, G.~Cucurull, A.~Casanova, A.~Romero, P.~Li{\`o}, and
  Y.~Bengio, ``Graph attention networks,'' in \emph{International Conference on
  Learning Representations}, 2018.

\bibitem{GuGao2019}
W.~Gu, F.~Gao, X.~Lou, and J.~Zhang, ``Link prediction via graph attention
  network,'' \emph{arXiv preprint arXiv:1910.04807}, 2019.

\bibitem{tang2020}
C.~Tang, J.~Sun, Y.~Sun, M.~Peng, and N.~Gan, ``A general traffic flow
  prediction approach based on spatial-temporal graph attention,'' \emph{IEEE
  Access}, vol.~8, pp. 153\,731--153\,741, 2020.

\bibitem{zhou2021}
H.~Zhou, D.~Ren, H.~Xia, M.~Fan, X.~Yang, and H.~Huang, ``{AST-GNN}: An
  attention-based spatio-temporal graph neural network for interaction-aware
  pedestrian trajectory prediction,'' \emph{Neurocomputing}, vol. 445, pp.
  298--308, 2021.

\bibitem{zhang2020}
Z.~Zhang, J.~Huang, and Q.~Tan, ``{SR-HGAT}: Symmetric relations based
  heterogeneous graph attention network,'' \emph{IEEE Access}, vol.~8, pp.
  165\,631--165\,645, 2020.

\bibitem{vaswani2017attention}
A.~Vaswani, N.~Shazeer, N.~Parmar, J.~Uszkoreit, L.~Jones, A.~N. Gomez,
  {\L}.~Kaiser, and I.~Polosukhin, ``Attention is all you need,'' in
  \emph{Advances in neural information processing systems}, 2017, pp.
  5998--6008.

\bibitem{2011Handbook}
R.~H. Hammack, W.~Imrich, S.~Klav{\v{z}}ar, W.~Imrich, and S.~Klav{\v{z}}ar,
  \emph{Handbook of product graphs}.\hskip 1em plus 0.5em minus 0.4em\relax CRC
  press Boca Raton, 2011, vol.~2.

\bibitem{Hammond2011}
D.~K. Hammond, P.~Vandergheynst, and R.~Gribonval, ``Wavelets on graphs via
  spectral graph theory,'' \emph{Applied and Computational Harmonic Analysis},
  vol.~30, no.~2, pp. 129--150, 2011.

\bibitem{Shuman2015}
D.~I. Shuman, C.~Wiesmeyr, N.~Holighaus, and P.~Vandergheynst,
  ``Spectrum-adapted tight graph wavelet and vertex-frequency frames,''
  \emph{IEEE Transactions on Signal Processing}, vol.~63, no.~16, pp.
  4223--4235, 2015.

\bibitem{Springfield1}
E.~Kuschel and B.~Strohl, ``{COVID-19 hospitalizations in the Ozarks near
  pre-delta variant numbers},''
  \url{https://www.ozarksfirst.com/local-news/local-news-local-news/covid-19-hospitalizations-in-the-ozarks-near-pre-delta-variant-numbers/},
  Sep. 15, 2021 [Online].

\bibitem{umass3}
K.~Wilkinson, ``{Over 100 breakthrough COVID-19 cases reported at UMass
  Amherst},''
  \url{https://www.wwlp.com/news/local-news/hampshire-county/over-100-breakthrough-covid-19-cases-reported-at-umass-amherst/},
  Sep. 9, 2021 [Online].

\bibitem{holyoke}
M.~Rosenfield, ``{‘Preventable Tragedy'}: Committee issues report on {COVID}
  deaths at {H}olyoke soldiers' home,''
  \url{https://www.nbcboston.com/news/coronavirus/preventable-tragedy-mass-committee-issues-report-on-covid-deaths-at-holyoke-soldiers-home/2389122/},
  May. 24, 2021 [Online].

\bibitem{marlborough2}
{Marlborough-ma.gov}, ``Position available -- {COVID} ambassador -- {PT},
  temporary,''
  \url{https://www.marlborough-ma.gov/sites/g/files/vyhlif7576/f/uploads/aa20-24_boh_covid_ambassador_pt_temp_0.pdf},
  2020 [Online].

\bibitem{marlborough1}
------, ``Marlborough {COVID-19} city updates,''
  \url{https://www.marlborough-ma.gov/sites/g/files/vyhlif7576/f/news/marlborough_covid-19_8.20.pdf},
  Aug. 20, 2020 [Online].

\bibitem{Harvard}
{Harvard-ma.gov}, ``{COVID-19} pandemic: Next steps for worker safety as
  organizations open during {COVID-19},''
  \url{https://www.harvard-ma.gov/sites/g/files/vyhlif676/f/uploads/tnec_reopening_business_during_covid_october_2020_flyer.pdf},
  Oct., 2020 [Online].

\bibitem{Harvard3}
------, ``{Harvard BOH update on local COVID-19 vaccination clinics},''
  \url{https://www.harvard-ma.gov/board-health/news/harvard-boh-update-local-covid-19-vaccination-clinics},
  Feb. 24, 2021 [Online].

\bibitem{news2020}
J.~Chesak, ``Which cities and states have had the best {COVID-19} response?''
  \url{https://www.verywellhealth.com/covid-19-response-city-state-response-5086171},
  Nov. 8, 2020 [Online].

\bibitem{massdata}
{Massachusetts Department of Public Health}, ``Archive of {COVID-19} cases in
  {M}assachusetts,''
  \url{https://www.mass.gov/info-details/archive-of-covid-19-cases-in-massachusetts},
  Nov., 2021 [Online].

\bibitem{toolforge}
{Geohack.toolforge.org}, \url{https://geohack.toolforge.org}, 2021 [Online].

\bibitem{vaishnav2016graph}
N.~Vaishnav and A.~Tatu, ``A graph downsampling technique based on graph
  fourier transform,'' \emph{arXiv preprint arXiv:1612.07542}, 2016.

\bibitem{nguyen2020wide}
T.~Nguyen, M.~Raghu, and S.~Kornblith, ``Do wide and deep networks learn the
  same things? uncovering how neural network representations vary with width
  and depth,'' \emph{arXiv preprint arXiv:2010.15327}, 2020.

\bibitem{Sandryhaila2014}
A.~Sandryhaila and J.~Moura, ``Big data analysis with signal processing on
  graphs: Representation and processing of massive data sets with irregular
  structure,'' \emph{IEEE Signal Processing Magazine}, vol.~31, no.~5, pp.
  80--90, 2014.

\bibitem{Valdivia2015}
P.~Valdivia, F.~Dias, F.~Petronetto, C.~T. Silva, and L.~G. Nonato,
  ``Wavelet-based visualization of time-varying data on graphs,'' in \emph{2015
  IEEE Conference on Visual Analytics Science and Technology (VAST)}.\hskip 1em
  plus 0.5em minus 0.4em\relax IEEE, 2015, pp. 1--8.

\bibitem{DalCol2017}
A.~{Dal Col}, P.~{Valdivia}, F.~{Petronetto}, F.~{Dias}, C.~T. {Silva}, and
  L.~G. {Nonato}, ``Wavelet-based visual analysis for data exploration,''
  \emph{Computing in Science Engineering}, vol.~19, no.~5, pp. 85--91, 2017.

\bibitem{DalCol2018}
A.~Dal~Col, P.~Valdivia, F.~Petronetto, F.~Dias, C.~T. Silva, and L.~G. Nonato,
  ``Wavelet-based visual analysis of dynamic networks,'' \emph{IEEE
  Transactions on Visualization and Computer Graphics}, vol.~24, no.~8, pp.
  2456--2469, 2018.

\bibitem{shuman2018distributed}
D.~I. Shuman, P.~Vandergheynst, D.~Kressner, and P.~Frossard, ``Distributed
  signal processing via chebyshev polynomial approximation,'' \emph{IEEE
  Transactions on Signal and Information Processing over Networks}, vol.~4,
  no.~4, pp. 736--751, 2018.

\bibitem{2017Introduction}
B.~A. Keen, ``Feature scaling with scikit-learn,''
  \url{https://benalexkeen.com/feature-scaling-with-scikit-learn/}, May. 10,
  2017 [Online].

\bibitem{yao2019developing}
H.~Yao, C.~Jiang, and Y.~Qian, \emph{Developing networks using artificial
  intelligence}.\hskip 1em plus 0.5em minus 0.4em\relax Springer, 2019.

\bibitem{umass1}
{UMass Amherst COVID-19 Informatiom}, ``All {UMass} athletic activities on
  pause,''
  \url{https://www.umass.edu/coronavirus/news/all-umass-athletic-activities-pause},
  Feb. 7, 2021 [Online].

\bibitem{news20210917}
W.~Katcher, ``In areas with highest {COVID} rates, {A}mherst leads
  {M}assachusetts communities over 10,000 people; see how your town compared,''
  \url{https://www.masslive.com/news/2021/09/in-areas-with-highest-covid-rates-amherst-leads-massachusetts-communities-over-10000-people-see-how-your-town-compared.html},
  Sep. 17, 2021 [Online].

\bibitem{umass2}
{UMass Amherst COVID-19 Information}, ``Campus {COVID-19} risk level raised
  from “elevated” to “high” risk,''
  \url{https://www.umass.edu/coronavirus/news/campus-covid-19-risk-level-raised-elevated-high-risk},
  Feb. 7, 2021 [Online].

\bibitem{amherstma1}
{AMHERSTMA.GOV - NEWS}, ``Town of {A}mherst {COVID-19} update - {S}eptember 10,
  2021,'' \url{https://www.amherstma.gov/CivicAlerts.aspx?AID=2966&ARC=5832},
  Sep. 10, 2021 [Online].

\bibitem{marlborough3}
{Marlborough-ma.gov}, ``Board of health meeting agenda,''
  \url{https://www.marlborough-ma.gov/board-health/agenda/board-health-meeting-agenda-5},
  Nov. 9, 2020 [Online].

\end{thebibliography}

%
%

\newpage

 





\end{document}